\documentclass[prd,nofootinbib,onecolumn,preprint,11pt]{revtex4-2}

\usepackage{graphicx}
\usepackage{bm}
\usepackage{float}
\usepackage{subcaption}
\usepackage{amsmath}
\usepackage{amsfonts}
\usepackage{color}
\usepackage{slashed}
\usepackage{fancyhdr}
\usepackage{mathrsfs}
\usepackage[colorlinks=true,linkcolor=black,citecolor=blue,filecolor=black,urlcolor=black,unicode]{hyperref}

\newcommand{\GN}{G_{\rm N}}
\newcommand{\mpl}{m_{\rm pl}}
\newcommand{\phib}{\overline{\varphi}}
\newcommand{\order}[1]{\mathcal{O}(#1)}
\newcommand{\abs}[1]{\lvert #1 \rvert}
\newcommand{\chubble}{\mathcal{H}}
\newcommand{\ttpart}{\text{tt}}

\newcommand{\dd}{{\rm d}}

\begin{document}
\title{Scalar Gravitational Waves Can Be Generated Even Without Direct Coupling Between Dark Energy and Ordinary Matter} 
\author{Li-Ying Chou$^1$, Yi-Zen Chu$^{1,2}$ and Yen-Wei Liu$^1$}
\affiliation{
$\,^1$Department of Physics, National Central University, Chungli 32001, Taiwan \\
$\,^2$Center for High Energy and High Field Physics (CHiP), National Central University, Chungli 32001, Taiwan}

\begin{abstract}
    \noindent We point out, the scalar sector of gravitational perturbations may be excited by an isolated astrophysical system immersed in a universe whose accelerated expansion is not due to the cosmological constant, but due to extra field degrees of freedom. This is true even if the source of gravitational radiation did not couple directly to these additional fields. We illustrate this by considering a universe driven by a single canonical scalar field. By working within the gauge-invariant formalism, we solve for the electric components of the linearised Weyl tensor to demonstrate that both the gravitational massless spin-2 (transverse-traceless) tensor and the (Bardeen) scalar modes are generated by a generic astrophysical source. For concreteness, the Dark Energy scalar field is either released from rest, or allowed to asymptote to the minimum in a certain class of potentials; and we compute the traceless tidal forces induced by gravitational radiation from a hypothetical compact binary system residing in such a universe. Though their magnitudes are very small compared to the tensors', spin zero gravitational waves in such a canonical scalar driven universe are directly sensitive to both the Dark Energy equation of state and the eccentricity of the binary's orbit.
\end{abstract}
\maketitle
\newpage
\tableofcontents

\section{Introduction}

Since the discovery of the accelerated expansion of the universe, (astro)physicists have wondered if this phenomenon is due to the presence of a positive cosmological constant $\Lambda$ in Einstein's equations or to the existence of extra field degree(s) of freedom -- usually dubbed ``Dark Energy" -- that are `modifying' gravity at astrophysical and cosmological distance scales. More recently, humanity has finally entered the era of gravitational wave signal driven astronomy and cosmology. It has therefore become increasingly important to ask: Could gravitational waves in a Dark Energy dominated universe contain direct evidence for its existence?

If the acceleration of the universe were due solely to a positive $\Lambda$, then the further it expands, the more its local geometry would approach de Sitter (dS) spacetime. As such, there has already been several theoretical studies -- see \cite{Ashtekar2015Asymptotics},\cite{Ashtekar2015Asymptoticsa},\cite{Ashtekar2015Asymptoticsb}, \cite{Chu2017Gravitational}, and \cite{Vega1999Generation} -- of the properties of gravitational waves propagating in a dS background. Not only has Einstein's equations sourced by an isolated hypothetical astrophysical system been linearized and solved; the corresponding gravitational quadrupole radiation formula has also been derived \cite{Ashtekar2015Asymptoticsb}. These works inform us, the only components of the gravitational perturbations of dS spacetime capable of carrying energy-momentum away from their material sources to cosmological distances is the massless spin$-2$ portion; i.e., what constitutes gravitational radiation in dS has the same oscillatory polarization modes as those in Minkowski spacetime. In this paper, we wish to assert that this spin$-2$ only character of gravitational waves no longer holds, if the universe were instead driven by extra field degree(s) of freedom and -- rather crucially -- even if ordinary matter did not couple directly to Dark Energy. The magnitude of this new channel of cosmological gravitational radiation is expected to be small relative to its spin-2 cousin when direct coupling is absent; but because it potentially provides us with direct evidence of Dark Energy, and because ``What constitutes a gravitational wave?" is a fundamental physics question-of-principle, we believe it deserves to be studied quantitatively.

We shall slowly inflate the universe -- for technical simplicity -- with a single canonical scalar field, so that the background spacetime is a de Sitter like one. We then proceed to solve for the linear gravitational perturbations emitted by an isolated astrophysical system residing in it. To this end, the gauge-invariant formalism shall be employed, so that the ensuing wave solutions are automatically invariant under infinitesimal coordinate transformations, thereby allowing without ambiguity their proper identification as `scalar' versus `tensor' sectors of gravitational radiation. Moreover, since the conformally invariant Weyl tensor is zero when evaluated on the background de Sitter like, and hence, Friedmann-Lema\^{i}tre-Robertson-Walker (FLRW) geometry, its first order perturbed cousin must be gauge-invariant. This motivates us to employ the gauge-invariant wave solutions to construct the electric part of the linearized Weyl tensor, in the frame of the observer at rest with the unperturbed universe. This physical computation yields the (irreducible) traceless part of the tidal forces acting on a small body due to the passage of the gravitational wave train. In turn, the oscillatory polarizations corresponding to the different patterns of squeezing and stretching can be extracted. Finally, we specialize to the compact binary system, the primary source of gravitational waves to date.

This paper is organised as follows. The basic setup of background spacetime and perturbation is introduced in \autoref{sec: background}. In \autoref{sec: gauge-invariant perturbation}, we apply the gauge-invariant perturbation theory \cite{Bardeen1980Gauge} and use this approach to obtain the equation of motion for perturbations. The equations of motion for propagating degrees of freedom are solved. We focus on the electric part of Weyl curvature which is associated with the tidal force in the \autoref{sec: Weyl tensor} and use the compact binary system to provide a concrete realisation of gravitational wave generation in \autoref{sec: binary}. Finally, we summarise our results and conclude in \autoref{sec: conclusion}. The flat spacetime metric is the mostly minus $\eta_{\mu\nu} = \text{diag}[1,-1,-1,-1]$; and natural units $\hbar = c = 1$ are used throughout this paper.

\section{Setup and Background Dynamics}
\label{sec: background}

The classical dynamics of our setup follows from extremizing the sum of the Einstein-Hilbert $\mathcal{S}_{\text{EH}}$, canonical scalar field $S_\varphi$ (``Dark Energy"), and an isolated hypothetical astrophysical system $S_\text{astro}$ actions with respect to all relevant field and matter degrees of freedom:
\begin{equation}
\mathcal{S}_\text{total} 
\equiv \mathcal{S}_{\text{EH}} + \mathcal{S}_\varphi + \mathcal{S}_{\text{astro}} ;
\end{equation}
where, if $\GN$ is Newton's constant, $\mpl \equiv (8\pi \GN)^{-1/2}$, and $\mathcal{R}$ is the Ricci scalar, 
\begin{align}
\mathcal{S}_{\text{EH}} &= - \frac{\mpl^2}{2} \int \dd^4x\;\sqrt{-g} \mathcal{R} , \\
\mathcal{S}_\varphi 	&=  \int \dd^4x\; \sqrt{-g} \left( \frac{1}{2} \nabla^\mu \varphi \nabla_\mu \varphi - V[\varphi] \right) .
\end{align}
The resulting equations of motion are, respectively, Einstein's
\begin{align}
\Delta_{\mu\nu}
\equiv G_{\mu\nu} - \mpl^{-2} T^{(\varphi)}_{\mu\nu} = \mpl^{-2} T^{(a)}_{\mu\nu} 
\end{align}
and the relativistic ``acceleration equals negative gradient of potential"
\begin{equation}
\Box \varphi + V'[\varphi]  = 0 .
\end{equation}
Here, $G_{\mu\nu}$ is Einstein's tensor; while $T^{(\varphi)}_{\mu\nu}$ and $T^{(a)}_{\mu\nu}$ are the energy-momentum-shear-stress tensor of Dark Energy and the astrophysical system respectively. We shall remain, for the moment, agnostic about the internal dynamics of the astrophysical system.
 
Exploiting conformal time $\eta$ and Cartesian spatial coordinates $\{x^i\}$, we shall assume the zeroth order solution to the Dark Energy scalar is spatially homogeneous $\overline{\varphi}[\eta]$; and we shall further treat the astrophysical system as a perturbation. This ensures the zeroth order `background' dynamics of the geometry is the FLRW universe $\bar{g}_{\mu\nu} = a[\eta]^2 \eta_{\mu\nu}$, while including the effects of $S_\text{astro}$ generates a metric perturbation we shall denote as $a^2 \chi_{\mu\nu}$. The full geometry is therefore
\begin{align}
\label{PerturbedFLRW}
g_{\mu\nu}[\eta,\vec{x}] = a^2[\eta] \left( \eta_{\mu\nu} + \chi_{\mu\nu}[\eta,\vec{x}] \right) .
\end{align}
Whereas, the full scalar field is
\begin{align}
\varphi[\eta,\vec{x}] \equiv \overline{\varphi}[\eta] + \psi[\eta,\vec{x}] ,
\end{align}
where $\psi[\eta,\vec{x}]$, like its gravitational counterpart $\chi_{\mu\nu}$, is considered to be a first order perturbation. For technical convenience, we shall place $\vec{x}=\vec{0}$ within the astrophysical system (say, at its center-of-mass). This way, $r \equiv |\vec{x}|$ would become the coordinate spatial distance between the astrophysical source to the observer at $(\eta,\vec{x})$ in the far zone, where the gravitational wave detector presumably lies.

Notationally, let us use $\delta_{n} (\dots)$ to denote the quantity in the parenthesis containing {\it all} terms with $a$ powers of $\chi_{\mu\nu}$ and $b$ powers $\psi$ such that $n = a+b$; so, for instance, $\delta_1 G_{\mu\nu}$ is the Einstein tensor linearized off the FLRW background, $\delta_1 T^{(\varphi)}_{\mu\nu}$ is the scalar $\varphi$'s stress tensor terms containing precisely one power of $\chi_{\mu\nu}$ and of $\psi$, and $\delta_1 T^{(a)}_{\mu\nu}$ is that of the astrophysical system containing exactly one power of $\chi_{\mu\nu}$. We shall witness below, both $\chi_{\mu\nu}$ and $\psi$ will be generated upon the inclusion of $S_\text{astro}$ in the dynamics. For the rest of this section, however, we shall focus on the background dynamics of $a[\eta]$ and $\overline{\varphi}[\eta]$, where $S_\text{astro}=0$ for now.

{\bf Low Energy Slow-Roll Inflation} \qquad The primary goal of the background Einstein-Dark Energy dynamics is to mimic a universe with equation-of-state $w$ very close to $-1$, since cosmological observations indicate that is indeed the case of our universe. In our setup, the pressure of Dark Energy is $P =  \frac12 a^{-2} \dot{\overline\varphi}^2 - V$, its energy density is $\rho = \frac12 a^{-2} \dot{\overline\varphi}^2 + V$, with each over-dot denoting a $\eta-$derivative; and therefore their ratio is
\begin{equation}
\label{EquationOfState}
w[\eta]
= -1 + \frac{\Dot{\overline{\varphi}}^2 / (a^2 V[\overline{\varphi}])}{1 + \Dot{\overline{\varphi}}^2 / (2 a^2 V[\overline{\varphi}])} 
\equiv -1 + \delta w[\eta].
\end{equation}
To ensure $\delta w \ll 1$, we therefore need $\Dot{\overline{\varphi}}^2 / (2 a^2 V[\overline{\varphi}]) \ll 1$. In this work, we shall achieve this by simply releasing the scalar field from rest, $\dot{\overline{\varphi}}[\eta_\star] = 0$, where $\eta_\star$ is some conformal time before the present time $\eta_0$. Or, in a class of potentials to be specified below, $\dot{\overline{\varphi}} \to 0$ while $\overline{\varphi}$ itself approaches the global minimum of its potential $V$. To yield a semi-realistic cosmology, we also need the reciprocal of the Hubble parameter to be roughly the age of the universe $a/\mathcal{H} \sim \mathcal{O}[14 \text{ Gyr}]$; namely, `low energy' inflation. Additionally, we shall impose the null energy condition $w \geq -1$, which implies $V$ needs to be non-negative.

Now, the homogeneity and isotropy of the background universe means the corresponding Einstein's equations
\begin{align}
\label{Einstein_0th}
\delta_0 \Delta_{\mu\nu}
= G_{\mu\nu}[\bar{g}] - \mpl^{-2} \overline{T}^{(\varphi)}_{\mu\nu} 
= 0
\end{align} 
are diagonal. Here, $G_{\mu\nu}[\bar{g}] \equiv \delta_0 G_{\mu\nu}$ is the Einstein tensor built solely out of the $\overline{g}_{\mu\nu} = a^2 \eta_{\mu\nu}$ and $\overline{T}^{(\varphi)}_{\mu\nu} \equiv \delta_0 T^{(\varphi)}_{\mu\nu}$ is the stress tensor of the Dark Energy scalar evaluated on $\overline{g}_{\mu\nu}$ and $\overline{\varphi}$. The $00$ component of eq. \eqref{Einstein_0th} reads, with $\mathcal{H} \equiv \dot{a}/a > 0$ for an expanding universe,
\begin{equation}
\label{Einstein_0th_00}
\mathcal{H}^2 = \frac{1}{3\mpl^2} \bigg( \frac{1}{2} \Dot{\overline{\varphi}}^2 + a^2 V[\overline{\varphi}] \bigg) .
\end{equation}
The diagonal spatial components are
\begin{align}
\label{Einstein_0th_ii}
- \dot{\mathcal{H}} - \frac{1}{2 }\mathcal{H}^2
= \frac{1}{2 \mpl^2} \left( \frac{1}{2} \dot{\overline{\varphi}}^2 - a^2 V \right) ;
\end{align}
or, equivalently, by taking into account the relation between $\dot{\overline{\varphi}}$, $V$, and $\delta w$ in eq. \eqref{EquationOfState},
\begin{equation}\label{EOM background acceleration}
\dot{\mathcal{H}} = - \frac{1}{2} (1+3w) \mathcal{H}^2 .
\end{equation}
Finally, the background Dark Energy scalar obeys
\begin{equation}
	\label{EOM of background scalar}
	\Ddot{\phib} + 2 \mathcal{H} \Dot{\phib} + a^2 V'[\phib]  = 0 .
\end{equation}

{\bf Release from rest} \qquad Since cosmological constraints are often phrased in terms of bounds on $w \equiv -1 + \delta w$, we now show that the scale factor $a[\eta]$ can in fact be entirely expressed in terms of $\delta w$. By taking the ratio of eq. \eqref{Einstein_0th_ii} to \eqref{Einstein_0th_00}, and without loss of generality (since $a>0$) parametrize
\begin{align}
a[\eta] = \exp\left[ \alpha[\eta] \right] ;
\end{align}
while taking into account eq. \eqref{EquationOfState}; we may readily derive an ordinary differential equation involving $\alpha$ and $\delta w$ only:
\begin{equation}
    \frac{\Ddot{\alpha}}{\dot{\alpha}^2} = - \frac{1}{2}(1+3w) .
\end{equation}
Let us choose to denote 
\begin{align}
	\eta_\star = -1/H_\star .
\end{align}
and proceed to impose the `initial conditions'
\begin{align}
\label{aInitialConditions}
a[\eta_\star] = 1
\qquad \text{and} \qquad
\dot{\alpha}[\eta_\star]^{-1} 
\equiv H_\star^{-1} = \sqrt{\frac{3}{8 \GN \pi V[\overline{\varphi}[\eta_\star]]}} \sim \mathcal{O}[14\text{ Gyr}] .
\end{align}
We then arrive at
\begin{equation}
\label{aInTermsOfdeltaw}
a[\eta] 
= \exp\bigg[ \int_{- H^{-1}_{\star}}^{\eta} \dd \eta''\; \bigg(  \frac{3}{2} \int_{- H^{-1}_{\rm \star}}^{\eta''} \delta w[\eta'] \;\dd\eta' - \eta''\bigg)^{-1} \bigg].
\end{equation}
Let us justify this un-conventional initial normalization of the scale factor. If we release the scalar field from rest,
\begin{align}
\label{ReleaseFromRest}
\dot{\overline{\varphi}}[\eta_\star]=0=\delta w[\eta_\star],
\end{align}
then we may suppose, for a suitably flat potential $V$, that $\dot{\overline{\varphi}}[\eta_\star]$ and $\delta w$ would remain small enough such that we may expand the integrand in powers of $\delta w$. This allows us express the scale factor as a deviation away from the de Sitter one
\begin{align}
	a_\text{dS}[\eta] \equiv -1/(H_\star \eta)
\end{align}
via the relation
\begin{equation}
\label{aDeviationFromdS}
a[\eta]
= a_\text{dS}[\eta] \cdot \exp\bigg[ -\frac{3}{2} \int_{-H^{-1}_{\star}}^{\eta} \frac{\dd \eta''}{\eta''^2} \int_{- H^{-1}_{\star}}^{\eta''} \delta w[\eta'] \dd\eta' + \mathcal{O}[\delta w^2] \bigg].
\end{equation}
If we choose the initial conditions in equations \eqref{aInitialConditions} and \eqref{ReleaseFromRest}, we may immediately solve for the initial second derivatives $\ddot{\alpha}[\eta_\star]$ and $\ddot{\overline{\varphi}}[\eta_\star]$ in terms of $H_\star$ and $V'_0 \equiv V'[\overline{\varphi}[\eta_\star]]$ via equations \eqref{Einstein_0th_ii} and \eqref{EOM of background scalar}. By taking $\eta-$derivatives of these same equations, we may then solve the initial third and higher derivatives of $\alpha$ and $\overline{\varphi}$ by iteration. This allows a series solution of $\alpha$ and $\overline{\varphi}$ to be constructed, in powers of $\ln a_\text{dS}$. In this manner, we may express various quantities of use later in this paper in terms of such a power series. The perturbed equation of state, for instance, is
\begin{align}
\label{EoS_PT1}
\delta w 
= \frac{{V'_0}^2}{3 \mpl^2 H_\star^4} (\ln a_\text{dS})^2
\left( 1 - 3 \ln a_\text{dS} + \left(\frac{21}{4}-\frac{V''_0}{3 H_\star^2}\right) (\ln a_\text{dS})^2 + \mathcal{O}\left[ (\ln a_\text{dS})^3 \right] \right) ;
\end{align}
with $V^{(n \geq 1)}[\overline{\varphi}[\eta_\star]] \equiv V^{(n)}_0$. We also need the first and second derivative terms:
\begin{align}
\label{EoS_PT2}
\frac{\delta \dot{w}}{\delta w}
&= \frac{2 H_\star \cdot a_\text{dS}}{\ln a_\text{dS}} \left( 1 -\frac{3}{2} \ln a_\text{dS} + \left(\frac{3}{4} - \frac{V''_0}{3 H_\star^2}\right) (\ln a_\text{dS})^2 + \mathcal{O}\left[ (\ln a_\text{dS})^4 \right] \right) ; \\
\label{EoS_PT3}
\frac{\delta \ddot{w}}{\delta w}
&= \frac{2 H_\star^2 a_\text{dS}^2}{\left(\ln a_\text{dS}\right)^2} \left( 1 - 5 \ln a_\text{dS} + \left(\frac{27}{4} - \frac{5V''_0}{3 H_\star^2}\right) (\ln a_\text{dS})^2 + \mathcal{O}\left[ (\ln a_\text{dS})^2 \right] \right) .
\end{align}
Whereas
\begin{align}
\label{calH_PT}
\mathcal{H}
= H_\star a_\text{dS} \cdot \left( 1 - \frac{4 \GN \pi {V'_0}^2}{3 H_\star^4} (\ln a_\text{dS})^3 + \mathcal{O}\left[ (\ln a_\text{dS})^4 \right]  \right) .
\end{align}
These solutions in equations \eqref{EoS_PT1}--\eqref{calH_PT} teach us that, for our Dark Energy driven universe to remain close to $w=-1$ after releasing the scalar field from rest, we ought to impose the low energy `slow roll' conditions
\begin{align}
\frac{{V'_0}^2}{\mpl^2 H_\star^4} \ll 1
\qquad \text{and} \qquad
\frac{V''_0}{H_\star^2} \ll 1 .
\end{align}
For technical reasons to be elaborated further below, we shall also assume that the second and higher time derivatives of the mass quadrupole moments are non-zero -- gravitational radiation production is active -- only strictly after $\eta_\star$.

{\bf Dynamical system analysis} \qquad Other than releasing the background Dark Energy scalar field from rest, what other circumstances would yield $0 \leq \delta w \ll 1$? Let us employ dynamical systems analysis to probe this question. As a start, let us choose $N\equiv \ln a$ (not to be confused with the $\ln a_\text{dS}$ above) as the evolution parameter and define the dimensionless variables
\begin{equation}
    x = \frac{\dot{\overline{\varphi}}}{\sqrt{6}\mpl \mathcal{H}}, \qquad 
    y = \frac{a \sqrt{V}}{\sqrt{3}\mpl \mathcal{H}}, \qquad 
    \lambda = \frac{a^2 V'[\overline{\varphi}]}{\sqrt{6}\mpl \mathcal{H}^2} .
\end{equation}
Then the first Friedmann equation in eq. \eqref{Einstein_0th_00} becomes the equation for a half cylinder, where
\begin{equation}\label{Friedmann eq dynamical sys}
    x^2 + y^2 = 1
    \qquad\text{and}\qquad
    y > 0 .
\end{equation}
This means $y = \sqrt{1-x^2}$, and we only need the equations for $x$ and $\lambda$. Using equations \eqref{Einstein_0th_ii} and \eqref{EOM of background scalar},
\begin{align}
	\label{dynamical sys 1}
	\frac{\dd x}{\dd N} 
	&= 3x^3 - 3x -\lambda \\
	\label{dynamical sys 3}
	\frac{\dd\lambda}{\dd N} 
	&= 6x^2 \lambda  + F \cdot x ;
\end{align}
Furthermore, the equation of state is
\begin{equation}
\label{EquationOfState_x}
w = -1 + 2x^2 .
\end{equation}
where
\begin{equation}
F = \frac{a^2 V''[\overline{\varphi}]}{\mathcal{H}^2}.
\end{equation} 
To make further progress, let us specialise a class of specific potentials which lead to
\begin{equation}\label{non-isolated fixed point}
F[x,\lambda] = 2k \frac{\lambda^2}{1-x^2} ,
\end{equation}
where $k$ is a model-dependent constant. This class of potentials contains positive and negative power laws as well as exponentials. For instance, the power-law potential given by 
\begin{equation}
	V[\phib] = \frac{g_n}{n!} \phib^n,
\end{equation}
for $n$ even, takes the form in eq. \eqref{non-isolated fixed point} with 
\begin{equation}
	k = \frac{n-1}{n} . 
\end{equation}
In practice, we remained agnostic about the specific form of $V$ but simply choose different values of $k$ for $F$ in eq. \eqref{non-isolated fixed point}, and proceed to numerically evolve equations \eqref{dynamical sys 1} and \eqref{dynamical sys 3} on the computer.

\begin{figure}[H]
	\centering
	\begin{subfigure}[b]{0.3\textwidth}
		\centering
		\includegraphics[width=2in]{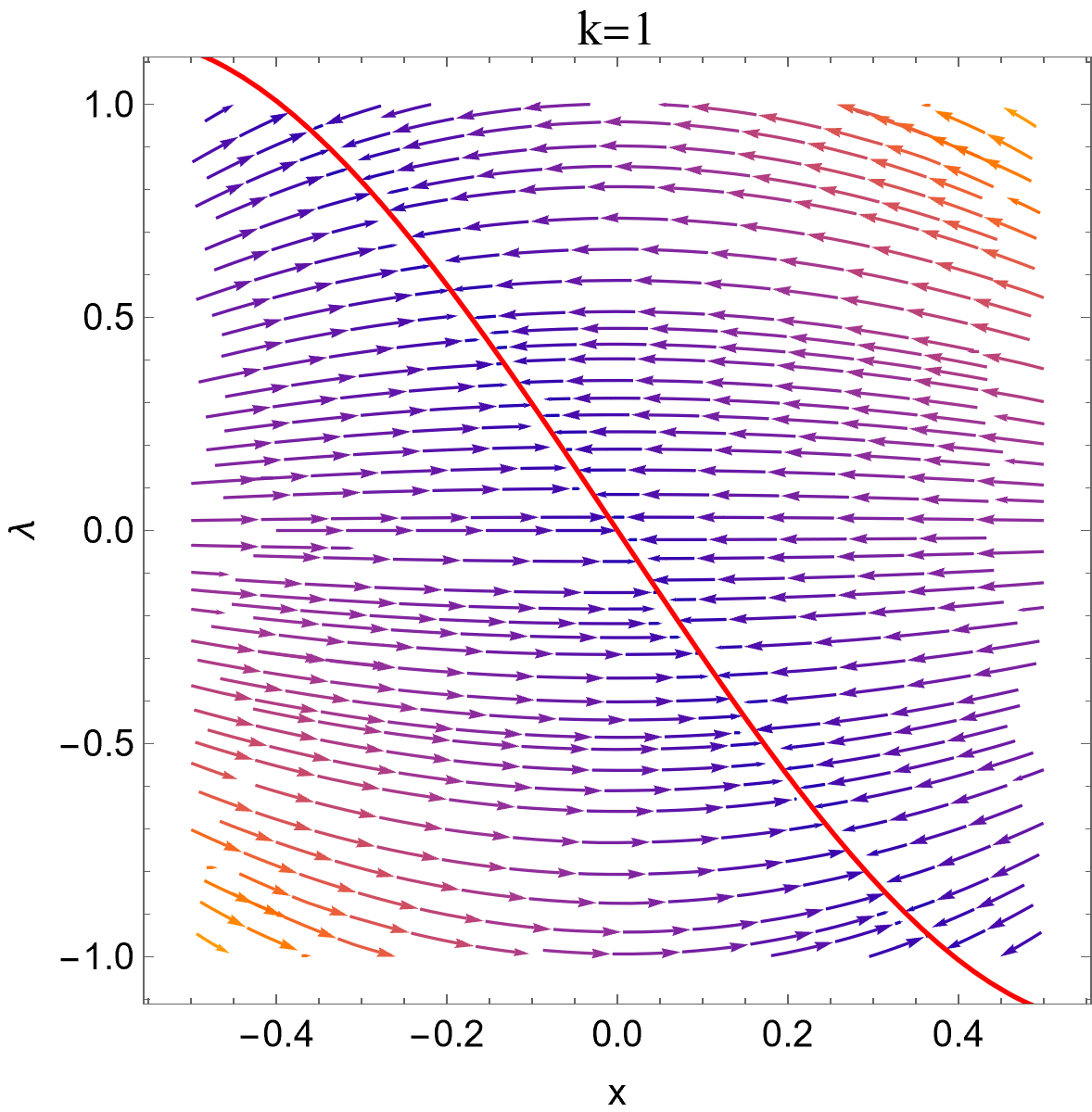}
	\end{subfigure}
	\begin{subfigure}[b]{0.3\textwidth}
		\centering
		\includegraphics[width=2in]{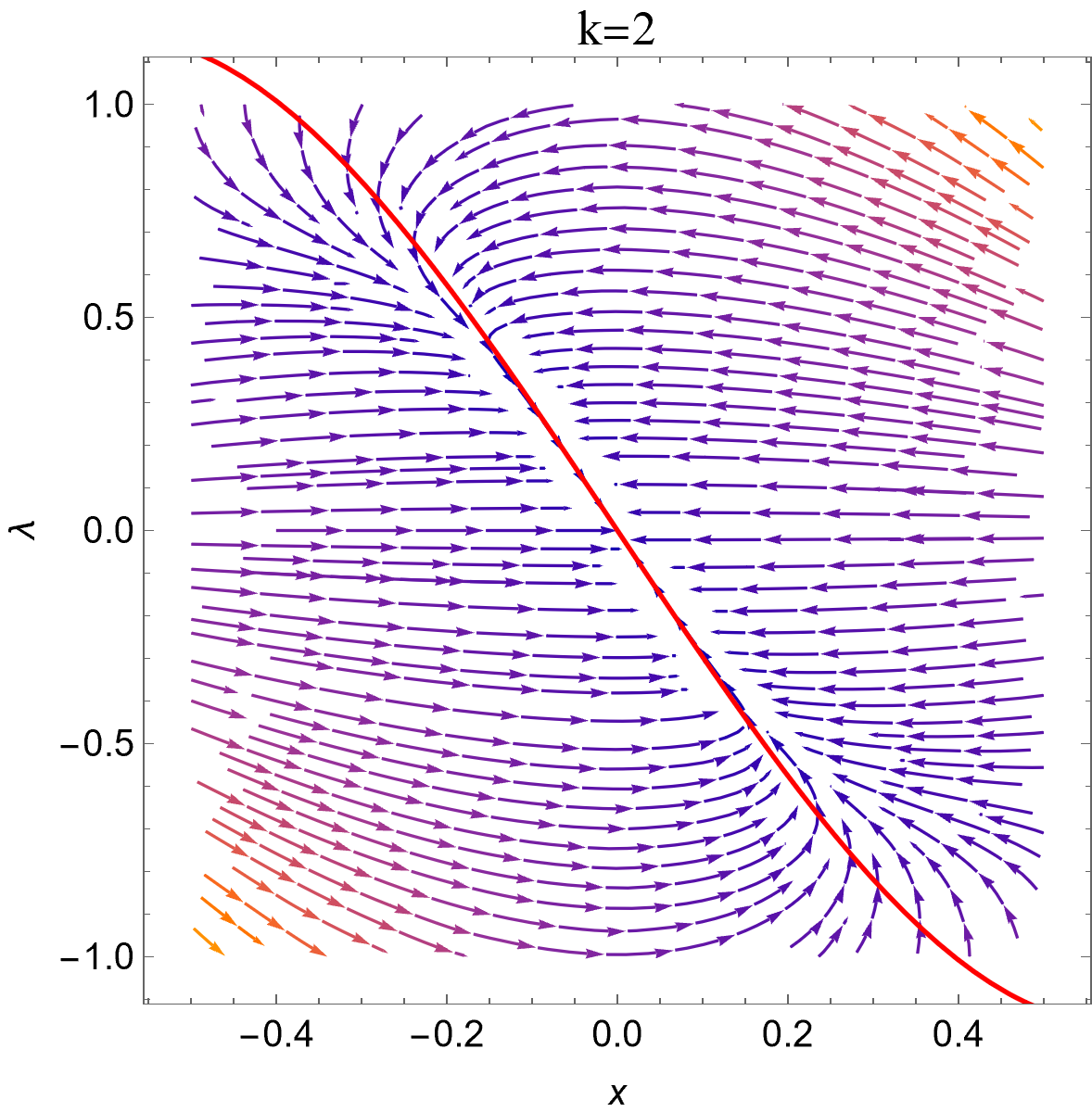}
	\end{subfigure}
	\begin{subfigure}[b]{0.3\textwidth}
		\centering
		\includegraphics[width=2in]{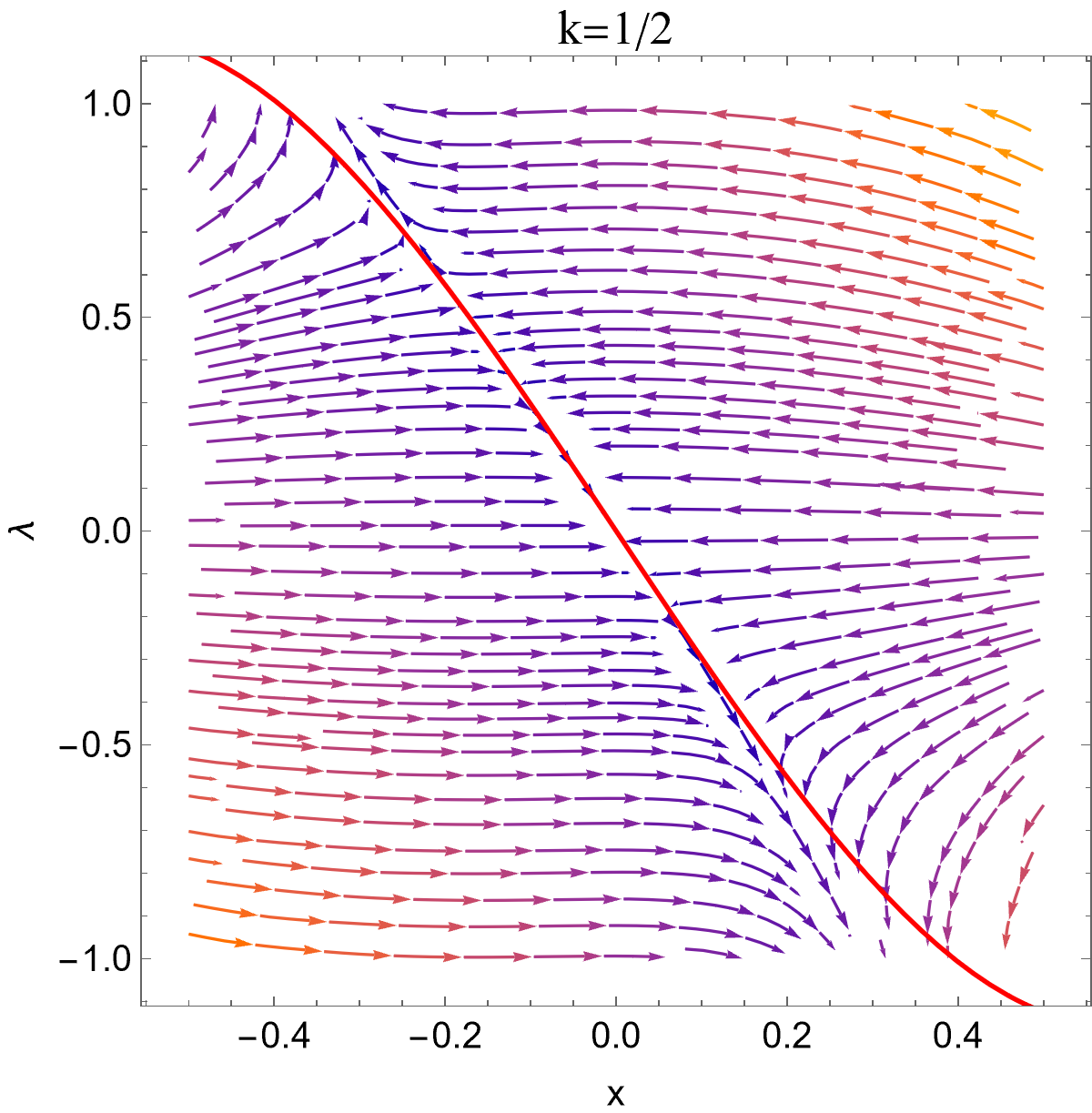}
	\end{subfigure}
	\caption{The numerical results for different values of $k$. The horizontal axis is $x$-axis, while the vertical axis is the $\lambda$-axis. The red curve is the $x$-nullcline given by $3x^3-3x = \lambda$. The figures above represent the system with $k=1$, $k=2$, and $k=1/2$ corresponding to exponential potential $e^{-\alpha\phib}$, reciprocal potential $1/\phib$, and quadratic potential $\phib^2$, respectively.}
	\label{fig:dynamical system}
\end{figure}

At this point, we observe that $(x,\lambda) = (0,0)$ is a fixed point, the sole solution to $(\dd x/\dd N, \dd \lambda/\dd N) = (0,0)$. Around $(0,0)$, eq. \eqref{EOM background acceleration} can be integrated directly to obtain the scale factor $a = a_{\rm dS}$. This asymptotic fixed point correspond to de Sitter spacetime. The Jacobian matrix at $(0,0)$ is given by 
\begin{equation}
J_0 =
    \begin{pmatrix}
         -3 & -1 \\ 
         0 	& 0
    \end{pmatrix} ;
\end{equation}
whose unit eigenvectors are $(1,0)^\text{T}$ and $(-1,3)^\text{T}/\sqrt{10}$ with respective eigenvalues $-3$ and $0$. This result means that the system evolves along $-x$ direction faster than along $(-1,3)^{\rm T}$. Such the fact suggests that the system first approaches to $x-$nucline and then is restricted on it.\footnote{In the literature of dynamical system, such a fixed point is called stable and non-isolated.} In Fig. \eqref{fig:dynamical system}, we plot the solutions $(x,\lambda)$ as trajectories on the 2D plane. We see that maintaining $0 \leq (\delta w = 2x^2) \ll 1$, eq. \eqref{EquationOfState_x} amounts to the restriction of the Dark Energy trajectory to lie on the thin vertical strip centered on the $\lambda-$axis.

%\begin{figure}[H]
 %   \centering
  %  \includegraphics[width=6.5in]{fig1.png}
  %  \caption{The numerical results for different values of $k$. The horizontal axis is $x$-axis, while the vertical axis is the $\lambda$-axis. The red curve is the $x$-nullcline given by $3x^3-3x = \lambda$. The figures above represent the system with $k=1$, $k=2$, and $k=1/2$ corresponding to exponential potential $e^{-\alpha\phib}$, reciprocal potential $1/\phib$, and quadratic potential $\phib^2$, respectively.}
   % \label{fig:dynamical system}
%\end{figure}

\begin{figure}[H]
	\centering
	\includegraphics[scale = 0.8]{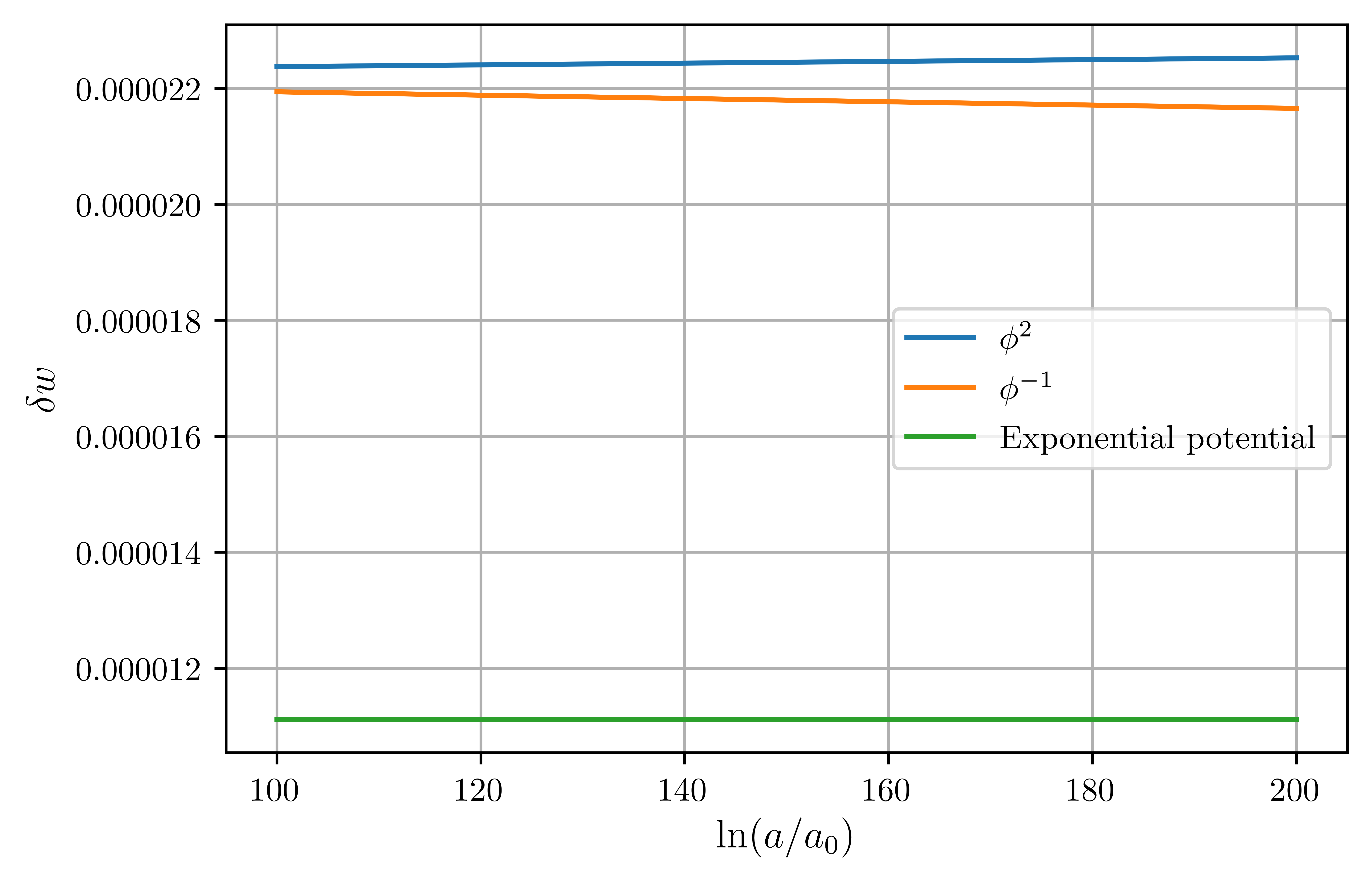}
	\caption{The numerical result of $\delta w$. The initial conditions are $x=\lambda=0.01$, and $a_0$ means the scale factor at initial time. We plot the time long enough after the initial time to study the asymptotic behaviour.}
	\label{fig:numerical dw}
\end{figure}

The evolution of $w$ can be studied in this framework. Taking derivatives of eq. \eqref{EquationOfState_x} while taking equations \eqref{dynamical sys 1} and \eqref{dynamical sys 3} into account, 
\begin{align}
\frac{\dd w}{\dd N} 
&= 4x (3x^3 - 3x - \lambda ),\label{first derivative of w} \\
\frac{\dd^2 w}{\dd N^2} 
&= 4\left( 36x^6 -54x^4 -21x^3\lambda + x^2(18-F)+ 9x\lambda^2 + \lambda^2 \right) .\label{second derivative of w}
\end{align}
The equation \eqref{first derivative of w} indicates that the first order derivative vanishes along the $x$-nullcline while the second order derivative \eqref{second derivative of w} is small along the $x$-nullcline as long as $\lambda\ll 1$. That is to say, $\delta w$ evolves relatively slowly comparing to $\mathcal{H}$. This fact is also demonstrated in the numerical results, which are presented in \autoref{fig:numerical dw}. According to the numerical results, it is clear that the change of $\delta w$ is small sufficiently long after initial time. 

\section{First Order Gauge-Invariant Perturbation Theory}
\label{sec: gauge-invariant perturbation}
\subsection{Equations}
By considering the hypothetical isolated astrophysical system as a first order perturbation, we now turn to solving Einstein's equations sourced by it, at the linear order in $\psi$ and $\chi_{\mu\nu}$:
\begin{align}
\label{Einstein_1stOrder}
\delta_1 \Delta_{\mu\nu}
= \delta_1 G_{\mu\nu} - \mpl^{-2} \delta_1 T^{(\varphi)}_{\mu\nu} 
&= \mpl^{-2} \overline{T}^{(a)}_{\mu\nu} , \\
\label{DarkEnergy_1stOrder}
\delta_1 \left( \Box \varphi + V'[\varphi] \right)
&= 0 .
\end{align}
{\bf Irreducible Decomposition} \qquad Because the zeroth order difference between the Einstein tensor and the dark energy scalar stress tensor is zero -- remember eq. \eqref{Einstein_0th} says $\delta_0 \Delta_{\mu\nu} = 0$ -- its first order version $\delta_1 \Delta_{\mu\nu}$ must be gauge-invariant: it remains the same under an infinitesimal change-of-coordinates (``gauge transformation") $x^\mu \to x^\mu + \xi^\mu$. Similar remarks apply to $\Box \varphi + V'[\varphi]$; its first order perturbation must too be gauge-invariant. We first decompose the metric perturbation into its irreducible components:
\begin{align}
	& \chi_{00} = E , \quad \chi_{0i} = \partial_i F + F_i , \nonumber \\
	& \chi_{ij} =  \chi^\text{TT}_{ij} + \partial_{\{i} D_{j\}} + \frac{D}{3}\delta_{ij} + \bigg( \partial_i \partial_j - \frac{1}{3} \delta_{ij} \vec{\nabla}^2\bigg) K ,
\end{align}
where the symmetrization $\{\dots\}$ is defined as, for e.g., $T_{\{ ij \}} = T_{ij} + T_{ji}$, $\vec{\nabla}^2 = \delta^{ij} \partial_i \partial_j$, and
\begin{equation}
\partial_i F_i 
= \delta^{ij} \chi^\text{TT}_{ij} 
= \partial_i \chi^\text{TT}_{ij} 
= \partial_i D_i = 0 .
\end{equation}
By studying how the above irreducible components transform under an infinitesimal change-of-coordinates, one may construct from them the gauge-invariant metric perturbation scalar, vector, and tensor variables
\begin{align}
\Psi 
&= E - \frac{2}{a} \partial_0 \left( a \left( F - \frac{\dot{K}}{2} \right) \right), \\
\Phi 
&= \frac{D - \vec{\nabla}^2 K}{3} + 2 \frac{\dot{a}}{a} \bigg( F - \frac{\dot{K}}{2} \bigg) , \\
V_i 
&= F_i - \dot{D}_i , \quad D_{ij} = \chi^\text{TT}_{ij} ;
\end{align}
as well as the one involving the Dark Energy perturbation,
\begin{equation}
\Pi = \psi + \dot{\overline{\varphi}} \bigg( \frac{\dot{K}}{2} - F \bigg) .
\end{equation}
We also need to decompose the $\chi_{\mu\nu}$-independent portion of the astrophysical system stress tensor:
\begin{align}
	\label{AstroStressTensor_000i}
	& \overline{T}^{(a)}_{00} = \varrho , \quad \overline{T}^{(a)}_{0i} = \Sigma_i + \partial_i \Sigma , \\
	\label{AstroStressTensor_ij}
	& \overline{T}^{(a)}_{ij} = \sigma_{ij} + \partial_{\{i} \sigma_{j\}} + \frac{\sigma}{3}\delta_{ij} + \bigg( \partial_i \partial_j - \frac{1}{3} \delta_{ij} \vec{\nabla}^2\bigg) \Upsilon ,
\end{align}
where
\begin{equation}
\partial_i \Sigma_i = \delta^{ij} \sigma_{ij} = \partial_i \sigma_{ij} = \partial_i \sigma_i = 0.
\end{equation}
In terms of these irreducible components, stress tensor conservation $\partial^\mu \overline{T}^{(a)}_{\mu\nu}=0$ at this order now reads
\begin{align}
	& \dot{\varrho} + 2 \frac{\dot{a}}{a}\varrho = \vec{\nabla}^2 \Sigma + \frac{\dot{a}}{a} (\varrho-\sigma) , \label{consevtion-a1} \\
	& \dot{\Sigma} + 2 \frac{\dot{a}}{a}\Sigma = \frac{\sigma}{3} + \frac{2}{3} \vec{\nabla}^2 \Upsilon , \label{consevtion-a2}\\
	& \dot{\Sigma}_i + 2 \frac{\dot{a}}{a}\Sigma_i = \vec{\nabla}^2 \sigma_i\label{consevtion-a3} .
\end{align}
{\bf First Order Perturbations} \qquad Next, we perform an irreducible decomposition on the linearized Einstein equations in eq. \eqref{Einstein_1stOrder}. Since the equations are themselves gauge-invariant, we expect them to be expressible solely in terms of the gauge invariant variables $\Phi$, $\Psi$, $V_i$, $D_{ij}$, and $\Pi$. In fact, extracting the scalar equations hands us
\begin{align}
	& \vec{\nabla}^2 \Phi - 3 \frac{\dot{a}}{a} \dot{\Phi} = \frac{1}{\mpl^2} \left(
	\dot{\overline{\varphi}} \dot{\Pi} + a^2 V'[\overline{\varphi}]\Pi + a^2V[\overline{\varphi}]\Psi + \varrho \right) \label{00}, \\
	& 3\Ddot{\Phi} + 6 \frac{\dot{a}}{a} \bigg(\dot{\Phi}+\frac{1}{2}\dot{\Psi}\bigg) - \vec{\nabla}^2 (\Phi-\Psi) = \frac{3}{\mpl^2} \left( \dot{\overline{\varphi}} \dot{\Pi} - a^2 V[\overline{\varphi}]\Psi - a^2 V'[\overline{\varphi}]\Pi + \frac{\sigma}{3} \right) \label{trace part}, \\
	& \dot{\Phi} + \frac{\dot{a}}{a} \Psi = \frac{1}{\mpl^2}(\dot{\overline{\varphi}} \Pi + \Sigma) \label{gradient part}, \\
	& \frac{1}{2} (\Phi - \Psi) = \frac{1}{\mpl^2} \Upsilon \label{tidal-shear} .
\end{align}
Next, the following manifestly gauge-invariant version of eq. \eqref{DarkEnergy_1stOrder} teaches us, $\Pi$ obeys a interacting wave equation -- with its interaction corresponding to the second derivative of the potential -- sourced by the gravitational scalars:\footnote{A similar situation occurs in constant equation-of-state cosmologies \cite{Chu2017More}, where it is the gravitational perturbations excited by the hypothetical astrophysical system that, in turn, sources the fluid perturbations.}
\begin{align}
a^2 \left(\overline{\Box} + V''[\overline{\varphi}]\right) \Pi = \frac{1}{2} \dot{\overline{\varphi}} ( \dot{\Psi}+3\dot{\Phi}) - a^2  V'[\overline{\varphi}] \Psi  \label{perturbed scalar field} ,
\end{align}
where $\overline{\Box}$ is the wave operator with respect to the background metric tensor $\Bar{g}_{\mu\nu} = a^2 \eta_{\mu\nu}$. The gauge-invariant vector equations from eq. \eqref{Einstein_1stOrder} are
\begin{align}
\frac{1}{2} \vec{\nabla}^2 V_i 
&= - \frac{1}{\mpl^2}\Sigma_i , \label{divergence-less part} \\
\frac{1}{2} \dot{V}_i + \frac{\dot{a}}{a} V_i 
&= \frac{1}{\mpl^2} \sigma_i \label{divergence-less part 2} .
\end{align}
It should be noted that, equations \eqref{divergence-less part} and \eqref{divergence-less part 2} are equivalent, as long as the conservation law in eq. \eqref{consevtion-a3} is respected.

Finally, the gauge-invariant tensor equation is
\begin{equation}\label{TT part}
	- \frac{a^2}{2} \overline{\Box} D_{ij} = \frac{1}{\mpl^2} \sigma_{ij} ,
\end{equation}
where $\overline{\Box}$ is the {\it scalar} wave operator with respect to $a^2 \eta_{\mu\nu}$.

{\bf Mixing is Inevitable} \qquad Before moving on to solve the gauge-invariant variables, let us observe that the scalar equations \eqref{00}--\eqref{gradient part} and \eqref{perturbed scalar field} involve a mixing of the metric perturbations $\Phi$ and $\Psi$ with the $\Pi$ associated with Dark Energy. This situation arises inevitably because equations \eqref{Einstein_1stOrder} and \eqref{DarkEnergy_1stOrder} receive contributions from both the gravitational and Dark Energy sectors, even though the full scalar $\varphi$ was not directly coupled to the astrophysical system. Since this mixing did not depend on the details of the Dark Energy stress tensor, we expect it to hold for generic scalar Dark Energy models. Moreover, we shall see below that, upon decoupling the scalar equations above, $\Phi$ and $\Psi$ obey wave equations. We believe this too shall continue to hold even if $\Pi$ were now associated with a Dark Energy model more complicated than the canonical one at hand.

\subsection{Solutions}

{\bf Mathematical Preliminaries} \qquad For the scalar and tensor equations below, we shall be faced with a wave equation of the form
\begin{align}
\label{FlatBoxPlusU}
\left(\partial^2 + U[\eta]\right) W[\eta,\vec{x}] = \mathcal{J} , 
\qquad
\partial^2 \equiv \eta^{\mu\nu} \partial_\mu \partial_\nu ;
\end{align}
where $U$ is some $\eta-$only `potential', $W$ is the wave itself, and $\mathcal{J}$ is some matter source. The retarded Green's function for $\partial^2 + U$ -- which obeys
\begin{align}
(\partial^2+U) \widehat{G}^+_U = \delta[\eta-\eta'] \delta^{(3)}[\vec{x}-\vec{x}']
\end{align}
-- can be shown \cite{Chu2015Transverse} to take the following form:
\begin{align}
\label{4DFLRWGreen}
\widehat{G}^+_U[\eta,\eta';R]
&\equiv \overline{G}^+_4[T,R] + \frac{\Theta[T-R]}{4\pi} \mathcal{G}_U^\text{(tail)}[\eta,\eta';R] , \\
\label{G4DFlat}
\overline{G}^+_4[T,R]
&\equiv \frac{\delta[T-R]}{4\pi R},
\qquad
T \equiv \eta-\eta', \qquad 
R \equiv |\vec{x}-\vec{x}'| ;
\end{align}
where the portion of the signal, proportional to the $\delta-$function, that travels on the forward null cone $T=R$ takes the flat spacetime massless Green's function form $\overline{G}^+_4$; while the inside-the-null cone piece, proportional to the step $(\Theta-)$function can be written as a derivative with respect to the flat spacetime world function
\begin{align}
\bar{\sigma} \equiv \frac{1}{2} T^2 - \frac{1}{2} R^2
\end{align}
of a certain homogeneous solution in 2D,
\begin{align}
\mathcal{G}_U^\text{(tail)}[\eta,\eta';R]
&= \frac{\partial \mathcal{G}_2[\eta,\eta';\bar{\sigma}]}{\partial \bar{{\sigma}}} .
\end{align}
This 2D homogeneous solution obeys, within the coordinate systems $(\eta,R)$ and $(\eta',R)$,
\begin{align}
\left( \partial_\eta^2 - \partial_R^2 + U[\eta] \right) \mathcal{G}_2[\eta,\eta';R]
= 0 =
\left( \partial_{\eta'}^2 - \partial_R^2 + U[\eta'] \right) \mathcal{G}_2[\eta,\eta';R] ;
\end{align}
with the null cone boundary condition
\begin{align}
\mathcal{G}_2[\eta,\eta'; \bar{\sigma} = 0] = 1 .
\end{align}
Additionally, if
\begin{align}
\label{Potential_kappa}
U[\eta] = - \frac{\kappa(\kappa+1)}{\eta^2}
\end{align}
for some constant $\kappa$, Nariai's ansatz \cite{Nariai1968Greens} allows for an explicit solution proportional to the derivative of the Legendre function $P_\kappa$ of order $\kappa$,
\begin{align}
\label{TailFunction}
\mathcal{G}_U^{(\text{tail})}
= \frac{1}{\eta \eta'} P'_\kappa\left[1+\frac{\bar{\sigma}}{\eta\eta'}\right].
\end{align}
{\bf Tensor} \qquad With these mathematical preliminaries in mind, let us first tackle the tensor case in eq. \eqref{TT part}. We may re-phrase it into the form in eq. \eqref{FlatBoxPlusU}:
\begin{equation}\label{TT wave eq 2}
\bigg( \partial^2 - \frac{\Ddot{a}}{a} \bigg) (aD_{ij}) = - \frac{2a}{\mpl^2} \sigma_{ij}.
\end{equation}
Recalling that eq. \eqref{aDeviationFromdS} may be argued to be an approximate solution to the scale factor in both cases where $\dot{\overline{\varphi}}[\eta_\star]=0$ and $a[\eta \to \eta_\star] \to a_\text{dS}[\eta]$, we may write the potential $\ddot{a}/a$
\begin{align}
	\frac{\Ddot{a}}{a} & = \frac{2}{\eta^2} - \frac{3\delta w}{2\eta^2} + \frac{6}{\eta^3} \int_{-H_\star^{-1}}^\eta \delta w[\eta'] \;\dd\eta' + \mathcal{O}[\delta w^2] \nonumber\\
	& \equiv \frac{2}{\eta^2} - \delta U[\eta].
\end{align}
If $\delta w$ were set to zero, the remaining $2/\eta^2$ term would yield the dS case, with $\kappa = -2$ or $\kappa = 1$; where -- referring to equations \eqref{4DFLRWGreen}, \eqref{Potential_kappa}, and \eqref{TailFunction} -- the retarded Green's function which obeys
\begin{align}
\left( \partial^2 - \frac{\ddot{a}_\text{dS}}{a_\text{dS}} \right) \widehat{G}^{+}_{\text{dS}_4} = \delta[\eta-\eta'] \delta^{(3)}[\vec{x}-\vec{x}'] ;
\end{align}
is
\begin{align}\label{GdS4}
\widehat{G}^{+}_{\text{dS}_4}[x,x']
&= \overline{G}^+_4[T,R] + \frac{\Theta[T-R]}{4\pi} H_\star^2 a_\text{dS}[\eta] a_\text{dS}[\eta'] .
\end{align}
The solution to eq. \eqref{TT wave eq 2} is thus, in the small $0 \leq \delta w \ll 1$ limit,
\begin{align}
\label{Tensor_Soln}
D_{ij}[\eta,\vec{x}] 
&= - \frac{2}{\mpl^2 a_\text{dS}[\eta]} \int \dd\eta' \int_{\mathbb{R}^3} \dd^3\vec{x}' a_\text{dS}[\eta'] \widehat{G}^{+}_{\text{dS}_4}[x,x'] \sigma_{ij}[\eta',\vec{x}']
+ \mathcal{O}[\delta w] .
\end{align}
Since the primary physical goal of this paper is the study of scalar gravitational waves, we shall not pursue the order $\delta w$ corrections to the tensor solution any further.

{\bf Bardeen Scalars} \qquad Turing to the scalars $\Phi$ and $\Psi$, let us proceed to decouple them from $\Pi$ in equations \eqref{00}--\eqref{gradient part}. First we eliminate $\Psi$ by inserting eq. \eqref{tidal-shear} into equations \eqref{00}--\eqref{gradient part}. Then, we subtract one-third of eq. \eqref{trace part} with \eqref{00}; followed by using eq. \eqref{gradient part} to eliminate $\Pi$ from the final result. This yields
\begin{equation}\label{Bardeen 1}
	\partial^2 \Phi + 2\bigg( \mathcal{H} - \frac{\Ddot{\overline{\varphi}}}{\dot{\overline{\varphi}}}\bigg) \dot{\Phi} + 2\bigg( \dot{\mathcal{H}} - \mathcal{H} \frac{\Ddot{\overline{\varphi}}}{\dot{\overline{\varphi}}} \bigg) \Phi = J(x)
\end{equation}
where
\begin{align}
	\label{Bardeen_Source}
	& J \equiv \frac{1}{\mpl^2} \left( \frac{\partial_0(a\Sigma)}{a} - \varrho + 2\mathcal{H} \Dot{\Upsilon} - 2\bigg( 2\mathcal{H} + \frac{\Ddot{\overline{\varphi}}}{\dot{\overline{\varphi}}} \bigg) \Sigma + 4 \bigg( \dot{\mathcal{H}} - \mathcal{H} \frac{\Ddot{\overline{\varphi}}}{\dot{\overline{\varphi}}} \bigg)\Upsilon \right) .
\end{align}
Now, eq. \eqref{Bardeen 1} may also be massaged into the massless scalar wave equation
\begin{equation}
	\frac{\dot{a}^3}{a^7 \dot{\overline{\varphi}}^2} \Box_\Omega \left(\frac{a^3}{\dot{a}} \Phi\right) = J[x] ,
\end{equation}
where the wave operator $\Box_\Omega$ is defined with respect to the fictitious geometry
\begin{equation}
g^{(\Omega)}_{\mu\nu}
\equiv \Omega^2 \eta_{\mu\nu} , \qquad 
\Omega \equiv \frac{\dot{a}}{a^2 \Dot{\overline{\varphi}}} .
\end{equation}
The geometry's conformally flat form indicates that the Bardeen scalar $\Phi$'s wavefront propagates at unit speed -- like its spin$-2$ $D_{ij}$ counterpart above -- with the null cone defined by $\bar{\sigma} = 0$.

Next, we transform eq. \eqref{Bardeen 1} into the form in eq. \eqref{FlatBoxPlusU}:
\begin{align}\label{eq:Phi}
\bigg( \partial^2 + U_\Phi\bigg) \left( \frac{a}{\Dot{\overline{\varphi}}} \Phi \right) = \frac{a}{\Dot{\overline{\varphi}}} J[x] , 
\end{align}
with the potential taking two equivalent forms
\begin{align}
\label{Potential_Phi_v1}
U_\Phi 
&= \dot{\mathcal{H}} - \mathcal{H}^2 + \frac{\dddot{\overline{\varphi}}}{\dot{\overline{\varphi}}} - 2 \bigg( \frac{\ddot{\overline{\varphi}}}{\dot{\overline{\varphi}}} \bigg)^2 \\
\label{Potential_Phi_v2}
&= \frac{3}{2} \delta w \cdot \mathcal{H}^2 + \frac{\delta\Dot{w}}{\delta w} \mathcal{H} + \frac{3}{4} \frac{\delta\Dot{w}^2}{\delta w^2} - \frac{1}{2} \frac{\delta\Ddot{w}}{\delta w} .
\end{align}

{\it Release from rest} \qquad We readily recognize, from equations \eqref{eq:Phi} and \eqref{Potential_Phi_v2}, that $\dot{\overline{\varphi}}[\eta_\star]=0$ is a singular limit. For technical simplicity, let us assume that the astrophysical source is actively producing gravitational radiation over a duration $[\eta_\text{i},\eta_\text{f}]$ that lies strictly {\it after} this dS-like transition; namely, $\eta_\text{i} > \eta_\star$. (In the non-relativistic limit, this amounts to assuming that $\partial_\eta^{n \geq 2} Q_{ij}[\eta<\eta_\text{i}]=0$, where $Q_{ij}$ is the mass quadrupole moments to be defined below.) If we release the scalar field from rest and recall the perturbative results in equations \eqref{EoS_PT2}--\eqref{calH_PT}, then $U_\Phi$ in eq. \eqref{Potential_Phi_v2} receives its most singular $\mathcal{O}\left[(\ln a_\text{dS})^{-2}\right]$ contributions from the two rightmost terms, with corrections that scale as $1/\ln a_\text{dS}$,
\begin{align}
U_\Phi 
= \frac{2}{(\eta-\eta_\star)^2} + \mathcal{O}\left[(\ln a_\text{dS})^{-1}\right] .
\end{align}
Comparing $U_\Phi$ with eq. \eqref{Potential_kappa} allows us to utilize equations \eqref{4DFLRWGreen} and \eqref{TailFunction} to deduce, the retarded Green's function satisfying
\begin{align}
\left( \partial_{\eta,\vec{x}}^2 + \frac{2}{(\eta-\eta_\star)^2} \right) \widehat{G}^+_4[x,x']
= \left( \partial_{\eta',\vec{x}'}^2 + \frac{2}{(\eta'-\eta_\star)^2} \right) \widehat{G}^+_4[x,x']
= \delta[\eta-\eta'] \delta^{(3)}[\vec{x}-\vec{x}'] 
\end{align}
is
\begin{align}
\label{BardeenGreensFunction_FromRest}
\widehat{G}^+_4[x,x'] 
&= \overline{G}^+_4[T,R] + \frac{\Theta[T-R]}{4\pi} \frac{\partial}{\partial \Bar{\sigma}} P_{\kappa_\pm}[1+s] , \\
\kappa_\pm &= (1/2)(-1 \pm i \sqrt{7}) ;
\end{align}
where $\overline{G}^+_4$ is the flat spacetime Green's function in eq. \eqref{G4DFlat} and $s = \Bar{\sigma}/((\eta-\eta_\star)(\eta'-\eta_\star))$. Note that $P_{\kappa_\pm}[1+s]$ is not only the same function, it is real despite the complex order $\kappa_\pm$; and, additionally, $P'_{\kappa_\pm}[1+s]$ itself begins at $-1$ when $s=0=\bar{\sigma}$ and asymptotes to zero as $s \to \infty$; namely, its total variation is $\mathcal{O}[1]$.

{\it Asymptotic behaviour}\qquad Now we consider the Bardeen scalar propagating in the asymptotic of background spacetime, which is associated with the behaviour of background along the $x$-nullcline and near the fixed point $(0,0)$. We consider the potential $V[\phib]$ such that $F$ takes the form \eqref{non-isolated fixed point}. Then the effective potential $U_\Phi$ evaluated on $x$-nullcline becomes
\begin{align}
U_\Phi 
&= -\frac{3}{2} \delta w \left( 5-6k + 3(k-1) \delta w \right) \chubble^2 .
\end{align}
From $\dd w/\dd N$ and $\dd^2 w/\dd N^2$ in \eqref{first derivative of w} and \eqref{second derivative of w} and the numerical results in fig. \eqref{fig:numerical dw}, we see that within certain classes of models and near the fixed point $(x,\lambda) \approx (0,0)$, $\delta w$ does not vary appreciably over many orders of magnitude change in the scale factor $a[\eta]$. Hence, here and below, we shall simply assume that $\delta w \equiv \delta w_0$ remains fixed from the epoch of gravitational radiation emission to its detection. The effective potential up to first order of $\delta w_0$ is thus
\begin{align}
 U_\Phi 
&\approx -\frac{3(5-6k)\delta w_0}{2\eta^2} + \mathcal{O}[\delta w_0^2]  .
\end{align}
(Recall that we have chosen $\eta_\star=0$.) Comparing $U_\Phi$ with eq. \eqref{Potential_kappa} allows us to invoke equations \eqref{4DFLRWGreen} and \eqref{TailFunction} to infer, the retarded Green's function satisfying
\begin{align}
\left( \partial_{\eta,\vec{x}}^2 - \frac{3(5-6k)\delta w_0}{2\eta^2} \right) \widehat{G}^+_4[x,x']
= \left( \partial_{\eta',\vec{x}'}^2 - \frac{3(5-6k)\delta w_0}{2\eta'^2} \right) \widehat{G}^+_4[x,x']
= \delta[\eta-\eta'] \delta^{(3)}[\vec{x}-\vec{x}'] 
\end{align}
is
\begin{align}
	\label{BardeenGreensFunction_Asymptotic}
	\widehat{G}^+_4[x,x'] 
	&= \overline{G}^+_4[T,R] + \frac{\Theta[T-R]}{4\pi} \frac{\partial}{\partial \Bar{\sigma}} P_{\kappa_\pm}[1+s] , \\
	\kappa_\pm &= \frac{1}{2} ( -1 \pm \sqrt{1+6(5-6k)\delta w_0}) ;
\end{align}
where, once again, $\overline{G}^+_4$ is defined in eq. \eqref{G4DFlat} and $P_{\kappa_\pm}[1+s]$ is the same function. Furthermore, as long as the value of $\delta w_0$ is small enough, $\kappa_\pm$ is a real number. The tail function can now be calculated perturbatively in powers of $\delta w_0$ by the integral representation
\begin{align}
P_\kappa[x] 
&= \frac{1}{\pi} \int_{-1}^1 \frac{\dd t}{\sqrt{1-t^2}} \big(x+t\sqrt{x^2-1}\big)^\kappa \\ 
&\approx 1 - \frac{3}{2} (5-6k)\delta w_0\ln\left[ \frac{2}{1+x} \right] 
+ \mathcal{O}\left[ \delta w_0^2 \right] .
\end{align}
Therefore,
\begin{equation}
\label{BardeenGreensFunction_AsymptoticApprox}
\frac{\partial}{\partial \Bar{\sigma}} P_{\kappa_\pm}[1+s] 
= \frac{3(5-6k)\delta w_0}{(\eta+\eta')^2-R^2} + \mathcal{O}\left[\delta w^2_0\right] .
\end{equation}
Since the universal $\overline{G}^+_4$ light cone term of $\widehat{G}_4^+$ in eq. \eqref{BardeenGreensFunction_Asymptotic} does not not contain any $\delta w$ dependence, we see that the tail term scales as $\delta w_0$ relative to it. 

{\it Summary} \qquad At this point, we may assert that
\begin{align}
	\label{Bardeen_Soln}
	\Phi[\eta,\vec{x}]
	= \frac{\Dot{\overline{\varphi}}[\eta]}{a[\eta]} \int\dd\eta' \int_{\mathbb{R}^3} \dd^3\vec{x}' \frac{a[\eta']}{\Dot{\overline{\varphi}}[\eta']} \widehat{G}_4^+[x,x'] J[x'] ; 
\end{align}
where, for the case where the Dark Energy scalar is released from rest, the Green's function $\widehat{G}_4^+[x,x']$ is given by eq. \eqref{BardeenGreensFunction_FromRest}; and for the asymptotic case it is given by eq. \eqref{BardeenGreensFunction_Asymptotic}, with the order $\delta w_0$ accurate tail given by eq. \eqref{BardeenGreensFunction_AsymptoticApprox}.

{\bf Vector} \qquad Finally, we solve the vector fields obeying eq. \eqref{divergence-less part},
\begin{align}
\label{Vector_Soln}
V_i[\eta,\vec{x}]
&= \frac{2}{\mpl^2} \int_{\mathbb{R}^3} \frac{\dd^3\vec{x}'}{4\pi|\vec{x}-\vec{x}'|} \Sigma_i[\eta,\vec{x}'] .
\end{align}
{\bf Acausality} \qquad We close this section with an important remark regarding the acausal character of these gauge invariant solutions. Firstly, since eq. \eqref{Vector_Soln} is a signal instantaneous in time, it manifestly does not obey relativistic causality. It turns out, so do its scalar and tensor cousins $\Phi$, $\Psi$ and $D_{ij}$. (Recall that $\frac{1}{2} (\Phi - \Psi) = \frac{1}{\mpl^2} \Upsilon$, so we only need to discuss $\Phi$.) The matter sources of $D_{ij}$ in eq. \eqref{Tensor_Soln} and $\Phi$ in eq. \eqref{Bardeen_Soln} (cf. eq. \eqref{Bardeen_Source}) are all spatially {\it nonlocal} functions of the astrophysical stress tensor, due to the projection process of equations \eqref{AstroStressTensor_000i} and \eqref{AstroStressTensor_ij} occurring in Fourier space -- see \S IV A of \cite{Chu2017More} for details. Due to this spatial `smearing out' of $\overline{T}^{(a)}_{\mu\nu}$, we expect -- like the constant$-w$ universe case \cite{Chu2020transverse} -- $D_{ij}[\eta,\vec{x}]$ and $\Phi[\eta,\vec{x}]$ (and, hence, $\Psi[\eta,\vec{x}]$) to be dependent on $\overline{T}^{(a)}_{\mu\nu}[\eta',\vec{x}']$ both inside and the outside of the light cone defined by $\eta-\eta' = |\vec{x}-\vec{x}'|$. To extract a physical and causal result, we now turn to computing from these gauge-invariant solutions the electric components of the Weyl tensor.

\section{Tidal Forces from Weyl Curvature}\label{sec: Weyl tensor}

\subsection{Physical Preliminaries}

If $U^\mu$ is tangent to the worldline of a freely-falling observer, and if an infinitesimal $\xi^\mu$ joins the observer's location to another nearby geodesic one, their relative tidal acceleration $a^\mu$ is driven by the Riemann tensor via the relation
\begin{align}
\label{TidalAcceleration}
a^\mu = - R^\mu_{\phantom{\mu}\nu \alpha\beta} U^\nu \xi^\alpha U^\beta .
\end{align}
Since the Weyl tensor is the traceless portion of Riemann, the corresponding irreducible component of the tidal force is
\begin{align}
\label{TraceFreeTidalForce}
C^\mu_{\phantom{\mu}\nu \alpha\beta} U^\nu \xi^\alpha U^\beta .
\end{align}
In a FLRW universe, the spacetime metric is conformally flat, which implies its Weyl tensor is zero. This, in turn, means that the linearized Weyl tensor $\delta_1 C^\mu_{\phantom{\mu}\nu \alpha\beta}$ is gauge-invariant. If we choose the geodesic to be the co-moving one,
\begin{align}
U^\mu = a^{-1} \delta^\mu_0 + \mathcal{O}\left[\chi_{\alpha\beta}\right] ,
\end{align}
the linearized version of eq. \eqref{TraceFreeTidalForce} now becomes
\begin{align}
\label{TraceFreeTidalForce_1stOrder}
a^{-2} \delta^\mu_i \delta_1 C^i_{\phantom{\mu}0 j 0} \xi^j .
\end{align}
This motivates us to insert the gauge-invariant $\Phi$, $\Psi$, $V_i$ and $D_{ij}$ obtained in the previous section to calculate the electric components
\begin{equation}
\label{LinearizedWeyl_Formula}
\delta_1 C^{i}_{\phantom{i}0j0} 
= \frac{1}{4} \left(  \Ddot{D}_{ij} + \vec{\nabla}^2 D_{ij} - \partial_{\{i} \Dot{V}_{j\}} + \bigg( \partial_i\partial_j - \frac{\delta_{ij}}{3} \vec{\nabla}^2\bigg) (\Phi + \Psi) \right) .
\end{equation}
We should mention that the Weyl tensor obeys a wave equation, which implies that it depends on the astrophysical stress tensor $\overline{T}^{(a)}_{\mu\nu}$ in a causal manner. As a consistency check of our calculations, we have verified that the acausal terms of $\Phi$, $\Psi$, $V_i$ and $D_{ij}$ in eq. \eqref{LinearizedWeyl_Formula} do indeed cancel. We then find  
\begin{align}
	\label{LinearizedWeyl_gPlusPhi}
    \delta_1 C^i_{\;\;0j0}(\eta,\Vec{x}) & = \delta_1 C^{(g)i}_{\;\;\quad 0j0}(\eta,\Vec{x}) + \delta_1 C^{(\Phi)i}_{\;\;\;\quad 0j0}(\eta,\Vec{x}) \nonumber\\
    & + 4\pi \GN\bigg\{ T^{(a)}_{ij}(\eta,\Vec{x}) - \frac{\delta_{ij}}{3} \bigg( T^{(a)}_{00}(\eta,\Vec{x}) + 2T^{(a)}_{\ell\ell}(\eta,\Vec{x}) \bigg)\bigg\}  ,
\end{align}
where $\delta_1 C^{(g)i}_{\;\;\quad 0j0}$ and $\delta_1 C^{(\Phi)i}_{\;\;\;\quad 0j0}$ are, respectively, the tensor $D_{ij}$ and Bardeen scalar $\Phi$ contributions. Unlike the Minkowski case, the trace-free tidal force in our Dark Energy universe receives contributions from both the scalar and the tensor sectors, providing theoretical evidence that scalar gravitational waves are indeed engendered even without direct coupling between $\varphi$ and ordinary matter. 

\subsection{Tensor Contributions}

Eq. \eqref{Tensor_Soln} tells us the tensor contribution to the linearized Weyl tensor in eq. \eqref{LinearizedWeyl_gPlusPhi} must coincide at zeroth order in $\delta w$ with the de Sitter one, which has been computed in Appendix A of \cite{Chu2021Gravitational} in arbitrary spacetime dimensions. Therefore,
{\allowdisplaybreaks\begin{align}
&\delta_1 C^{(g)i}_{\;\;\quad 0j0}[\eta,\Vec{x}]  
= \frac{1}{4} \bigg( \ddot{D}_{ij} + \vec{\nabla}^2 D_{ij}\bigg)_{\text{causal}} \nonumber\\
& =- 8\pi \GN \int_{\mathbb{R}^3} \dd^3 \vec{x}'\; \int \dd\eta' \;  \frac{a[\eta']}{a[\eta]} \bigg\{ \left( \ddot{\widehat{G}}^{+}_{\text{dS}_4} - \chubble(\eta) \dot{\widehat{G}}^{+}_{\text{dS}_4} - \dot{\chubble}(\eta) \widehat{G}^{+}_{\text{dS}_4} \right) \nonumber\\
&\qquad\qquad\qquad\qquad\qquad\qquad
\times \left( T^{(a)}_{ij}[\eta',\Vec{x}'] + \frac{1}{2}\delta_{ij}\bigg( T^{(a)}_{00}[\eta',\Vec{x}'] - T^{(a)}_{\ell\ell}[\eta',\Vec{x}'] \bigg) \right) \nonumber\\
&\qquad\qquad\qquad\qquad - \partial_0 \partial_{\{i} \overline{G}^+_4 T^{(a)}_{j\} 0}[\eta',\Vec{x}']
+\frac{1}{2} \delta_{ij} \chubble[\eta'] \partial_0 \overline{G}^+_4 \bigg( T^{(a)}_{00}[\eta',\Vec{x}'] + T^{(a)}_{\ell\ell}[\eta',\Vec{x}'] \bigg) \nonumber\\
&\qquad\qquad\qquad\qquad + \frac{1}{2} \partial_i\partial_j \overline{G}^+_4  \bigg( T^{(a)}_{00}[\eta',\Vec{x}'] + T^{(a)}_{\ell\ell}[\eta',\Vec{x}'] \bigg)\bigg\} + \mathcal{O}[\delta w] ;
\end{align}}
where $\widehat{G}^{+}_{\text{dS}_4}[x,x']$ is the dS Green's function in eq. \eqref{GdS4} and $\overline{G}^+_4$ is the flat spacetime one is eq. \eqref{G4DFlat}.

\textbf{Far Zone, Non-Relativistic Limits} \qquad For studying the response of a gravitational wave detector placed at astrophysical or cosmological distances from the material source, we may now take the far zone and non-relativistic limits. If $\tau_c$ and $r_c$ respectively denote the characteristic timescale and length scale of the astrophysical system, and $H \equiv \mathcal{H}/a$ the Hubble parameter, 
\begin{align}
    \delta_1 C^{(g|\gamma)i}_{\;\quad\quad 0j0}[\eta,\Vec{x}] 
    & = - \frac{2\GN}{r} \frac{a[\eta_r]}{a[\eta]} P_{ijab} \partial^2_0 \int \dd^3\vec{x}'\; T^{(a)}_{ab}[\eta_r,\Vec{x}']+ \mathcal{O}[H\tau_c,\tau_c/(ar),r_c/(ar)] ,
\end{align}
where $\eta_r \equiv t-r$ is retarded time, the projectors are defined as
\begin{equation}
\label{ttProjector}
P_{ijab} = \frac{1}{2} P_{a\{i}P_{j\}b} - \frac{1}{2} P_{ij}P_{ab}, \quad\quad P_{ij} = \delta_{ij} - \widehat{r}_i \widehat{r}_j ;
\end{equation}
and we have used the conservation law in non-relativistic limit
\begin{align}
    & \partial_j T^{(a)}_{0j} = \Dot{T}{}^{(a)}_{00} + \order{H\tau_c},\\
    & \partial_j T^{(a)}_{ij} = \Dot{T}{}^{(a)}_{i0} + \order{H\tau_c};
\end{align}
and the far zone identity $\partial_i \approx -\widehat{r}_i \partial_0$. Define the mass quadrupole moment as
\begin{equation}
\label{QuadrupoleMoment}
Q_{ij}[\eta] 
\equiv a^2[\eta] \int_{\mathbb{R}^3} \dd^3x\; x_i x_j T^{(a)}_{00}[\eta,\Vec{x}] .
\end{equation}
By conservation, we can obtain that 
\begin{equation}
	\label{LinearizedWeyl_TensorNR}
    \delta_1 C^{(g|\gamma)i}_{\;\quad\quad 0j0}[\eta,\Vec{x}] 
    \approx - \frac{\GN}{a[\eta]r} \frac{\partial^4_{\eta_r} Q^{(\ttpart)}_{ij}[\eta_r]}{a^2[\eta_r]} ,
\end{equation}
with the `transverse-traceless' quadrupole defined as $Q^{\text{(tt)}}_{ij} \equiv P_{ij ab} Q_{ab}$; with the $P_{ij ab}$ in eq. \eqref{ttProjector} ensuring
\begin{align}
\delta^{ij} Q^{(\ttpart)}_{ij} = 0 = Q^{(\ttpart)}_{ij} \widehat{r}^i .
\end{align}

\subsection{Bardeen Scalar Contributions}

Next, the Bardeen scalars' contribution to the linearized Weyl tensor in eq. \eqref{LinearizedWeyl_gPlusPhi} everywhere exterior to the astrophysical bodies (where $\overline{T}^{(a)}_{\mu\nu}[\eta,\vec{x}]=0$) is
{\allowdisplaybreaks\begin{align}
    & \delta_1 C^{(\Phi)i}_{\;\;\quad 0j0}[\eta,\Vec{x}] = \frac{1}{4} \left\{ \bigg( \partial_i\partial_j-\frac{\delta_{ij}}{3}\vec{\nabla}^2 \bigg) 2\Phi \right\}_{\text{causal}}\nonumber\\
    & =  -6\pi \GN \int_{\mathbb{R}^3} \dd^3\vec{x}' \int \dd\eta' \; \frac{a[\eta']}{a[\eta]} \mathcal{H}[\eta] \mathcal{H}[\eta'] \left( \delta_{ij} - 3 \widehat{R}_i \widehat{R}_j \right) \left( \frac{1}{3}\widehat{G}^+_4 - \frac{1}{R} \frac{\partial}{\partial R} \widehat{Q}^+_4 \right) \nonumber\\
    &\qquad\qquad \qquad\qquad \times\sqrt{\delta w[\eta] \delta w[\eta']} \left( T^{(a)}_{00}[\eta',\Vec{x}'] + T^{(a)}_{\ell\ell}[\eta',\Vec{x}'] \right) ,
    %& =  -6\pi \GN \int_{\mathbb{R}^3} \dd^3\vec{x}' \int \dd\eta' \; \frac{a[\eta']}{a[\eta]}\chubble[\eta] \chubble[\eta'] \bigg( \frac{\delta_{ij}}{3} \widehat{G}^+_4 - \partial_i\partial_j \widehat{Q}^+_4\bigg)\nonumber\\
    %& \qquad\quad\qquad\qquad\qquad\qquad \times\sqrt{\delta w[\eta] \delta w[\eta']}\bigg( T^{(a)}_{00}[\eta',\Vec{x}'] + T^{(a)}_{\ell\ell}[\eta',\Vec{x}'] \bigg).
\end{align}}
where $\widehat{R}^i \equiv (x^i-x'^i)/R$ and
\begin{align}
\widehat{Q}^+_4 
&= \int_{\eta'}^\eta \dd\eta_1 \frac{v[\eta]}{v[\eta_1]}\int_{\eta'}^{\eta_1} \dd\eta_2\;\frac{v[\eta_2]}{v[\eta_1]} \widehat{G}^+_4[\eta_2,\eta',R] , \\
\label{vDef}
v &\equiv \chubble/(a \Dot{\phib}) . 
\end{align}
To arrive at the above $\delta_{ij} - 3 \widehat{R}_i \widehat{R}_j$ tensor structure, corresponding to a scalar polarization pattern in the far zone \cite{Chu2021Gravitational}, we had employed the relation %and then obtain
\begin{equation}
    \partial_i\partial_j \widehat{Q}^+_4 = \widehat{R}_i \widehat{R}_j \widehat{G}^+_4 + (\delta_{ij} - 3  \widehat{R}_i \widehat{R}_j) \frac{1}{R} \frac{\partial}{\partial R} \widehat{Q}^+_4 ,
\end{equation}
which we, in turn, derived from the equations of motion obeyed by $\widehat{G}^+_4$. %where 

{\bf Far Zone, Non-Relativistic Limits} \qquad %Like spin-2 case, 
Let us now consider the linearized electric Weyl tensor in the far zone and non-relativistic limits. This involves performing a multi-pole expansion involving the mass $\overline{T}^{(a)}_{00}$ and sum-of-pressures $\overline{T}^{(a)}_{\ell\ell}$ densities:
\begin{align}
	& Q^{(n)}[\eta] = a^{n+1}[\eta] \int_{\mathbb{R}^3} \dd^3\vec{x}'\; (\Vec{x}'\cdot\widehat{r})^n T^{(a)}_{00}[\eta,\Vec{x}'], \\
	& P^{(n)}[\eta] = a^{n+1}[\eta] \int_{\mathbb{R}^3} \dd^3\vec{x}'\; (\Vec{x}'\cdot\widehat{r})^n T^{(a)}_{\ell\ell}[\eta,\Vec{x}'].
\end{align}
When $n=0$,
\begin{align}
Q^{(0)} \equiv M 
\qquad \text{and} \qquad
P^{(0)} \equiv P .
\end{align}
We shall also break up the linear electric Weyl tensor into the signal on the null cone $\delta_1 C^{(\Phi|\gamma)i}_{\;\quad\quad 0j0}$ and inside the null cone $\delta_1 C^{(\Phi|{\rm tail})i}_{\quad\qquad 0j0}$: 
\begin{align}
\delta_1 C^{(\Phi)i}_{\;\;\quad 0j0}[\eta,\Vec{x}] 
&= \delta_1 C^{(\Phi|\gamma)i}_{\;\quad\quad 0j0} + \delta_1 C^{(\Phi|{\rm tail})i}_{\quad\qquad 0j0} .
\end{align}
We should point out that the light cone term is contributed by the term involving $(1/3)\widehat{G}^+_4$ only. Up to order $\delta w$, the null cone signal is
\begin{align}
&\delta_1 C^{(\Phi|\gamma)i}_{\;\quad\quad 0j0}
= -\frac{\GN}{2a[\eta]r} (\delta_{ij}-3\widehat{r}_i\widehat{r}_j)\chubble[\eta]\chubble[\eta_r] \sqrt{\delta w[\eta]\delta w[\eta_r]} \left( M[\eta_r] + P[\eta_r] + \mathcal{O}[H\tau_c,r_c/(ar)] \right) 
\end{align}
Denoting the trace of pressure quadrupole by 
\begin{equation}
P_{ij} = \int \dd^3 x\; a^3[\eta] x_i x_j T^{(a)}_{\ell\ell}[\eta,\vec{x}] ,
\end{equation}
the sum of pressures may be related to the mass quadrupoles using conservation of stress-energy, yielding
\begin{equation}
P
= \frac{1}{2a}\partial_\eta \left( \frac{\partial_\eta Q_{\ell\ell}}{a} - \frac{\chubble}{a} (2 Q_{\ell\ell}-P_{\ell\ell}) \right) ,
\end{equation}
where the $Q_{ij}$ has already been defined in eq. \eqref{QuadrupoleMoment}. In the high frequency limit, where $\mathcal{H} \tau_c \ll 1$, the pressure monopole can be approximated as the acceleration of the trace of mass quadrupole,
\begin{equation}
\label{PLowestOrder}
P \approx \frac{\partial^2_\eta Q_{\ell\ell}}{2a^2} .
\end{equation}
Keeping only this dominant contribution to the pressure term,
\begin{align}
\label{LinearizedWeyl_NullFront}
\delta_1 C^{(\Phi|\gamma)i}_{\;\quad\quad 0j0}
&= -\frac{\GN}{2a[\eta]r} (\delta_{ij}-3\widehat{r}_i\widehat{r}_j)\chubble[\eta]\chubble[\eta_r] \sqrt{\delta w[\eta]\delta w[\eta_r]} \left( M[\eta_r] + \frac{\ddot{Q}_{\ell\ell}[\eta_r]}{2a^2[\eta_r]} 
+ \mathcal{O}\left[ H\tau_c,\frac{r_c}{ar}\right] \right) .
\end{align}
The tail signal of the linearized electric Weyl tensor is
\begin{align}
\label{LinearizedWeyl_Tail}
\delta_1 C^{(\Phi|{\rm tail})i}_{\quad\qquad 0j0} 
& = - \frac{3\GN}{2} \int_{\mathbb{R}^3} \dd^3\vec{x}' \int_{-\infty}^{\eta-R} \dd \eta' \; \frac{a[\eta']}{a[\eta]} \chubble[\eta]\chubble[\eta'] \sqrt{\delta w[\eta] \delta w[\eta']} (\delta_{ij}-3\widehat{R}_i\widehat{R}_j)\nonumber\\
& \qquad\qquad\qquad\qquad
\times\Delta[\eta,\eta',R] \left( T^{(a)}_{00}[\eta',\Vec{x}'] + T^{(a)}_{\ell\ell}[\eta',\Vec{x}'] \right) ,
\end{align}
where
\begin{align}
\Delta[\eta,\eta',R] 
&\equiv  \frac{1}{3}\frac{\partial P_{\kappa_\pm}[1+s]}{\partial \Bar{\sigma}} + \int_{\eta'+R}^\eta \dd\eta_1 \frac{v[\eta]}{v[\eta_1]} \int_{\eta'+R}^{\eta_1} \dd\eta_2 \; \frac{v[\eta_2]}{v[\eta_1]} \left. \frac{\partial^3}{\partial\Bar{\sigma}^3} P_{\kappa_\pm}\left[ 1+\frac{\bar{\sigma}}{\eta_2 \eta'} \right] \right\vert_{\bar{{\sigma}}=(1/2)(\eta_2-\eta')^2-(1/2)R^2} \nonumber\\
& + \frac{1}{R} \frac{\partial P_{\kappa_\pm}[1+s]}{\partial\Bar{\sigma}}\bigg|_{s=0}\int_{\eta'+R}^\eta \dd\eta_1 \frac{v[\eta]}{v[\eta_1]} \frac{v[\eta'+R]}{v[\eta_1]} + \frac{1}{R^2} \bigg(  \frac{v[\eta]}{v[\eta'+R]} - \int_{\eta'+R}^\eta \dd\eta_1\;  \frac{v[\eta]}{v[\eta_1]} \frac{\Dot{v}[\eta'+R]}{v[\eta_1]}\bigg)\nonumber\\
& + \frac{1}{R^3}\int_{\eta'+R}^\eta \dd\eta_1\;  \frac{v[\eta]}{v[\eta_1]} \frac{v[\eta'+R]}{v[\eta_1]}, \\
s&=\Bar{\sigma}/((\eta-\eta_\star)(\eta'-\eta_\star)) .
\end{align}
Multi-pole expansion of this tail signal involves $r-$derivatives of $\Delta[\eta,\eta',r]$; i.e., with $R \equiv |\vec{x}-\vec{x}'|$ replaced with $r \equiv |\vec{x}|$. For the case where the Dark Energy $\overline{\varphi}$ were released from rest, each higher derivative term would scale with one more power of $\abs{\Vec{x}'}/|\eta'-\eta_\star|$, which in turn goes as
\begin{equation}
\frac{\abs{\Vec{x}'}}{\abs{\eta'-\eta_\star}} 
\approx H[\eta'] a[\eta'] \abs{\Vec{x}'} \left(\frac{a[\eta']}{a[\eta_\star]}-1\right)^{-1} .
\end{equation}
Since $\vec{x}'$ runs over all spatial locations where $\overline{T}^{(a)}_{\mu\nu} \neq 0$, $|\vec{x}'|$ scales as the characteristic size of the system $r_c$; and, hence, $H[\eta'] a[\eta'] \abs{\Vec{x}'} \ll 1$. Although the remaining factor can be large if the duration of active production of gravitational waves is too close to the dS transition point -- i.e., if $\eta_\text{i} \to \eta_\star$ -- since our current setup is merely a toy model for Dark Energy, we shall assume that we are studying systems emitting late enough $\eta_\text{i} > \eta_\star$, so that $\abs{\Vec{x}'}/|\eta'-\eta_\star| \ll 1$. Taking eq. \eqref{PLowestOrder} into account,
\begin{align}
\label{LinearizedWeyl_Tail_FromRest}
& \delta_1 C^{(\Phi|{\rm tail})i}_{\quad\qquad 0j0}\nonumber\\
& \approx -\frac{3\GN \sqrt{\delta w[\eta]}}{2a[\eta]} (\delta_{ij}-3\widehat{r}_i\widehat{r}_j)  \int_{\eta_i}^{\eta_r} \dd\eta' \;\chubble[\eta] \chubble[\eta'] \sqrt{\delta w[\eta']} \Delta[\eta,\eta';r] \left( M[\eta'] + \frac{\ddot{Q}_{\ell\ell}[\eta']}{2a^2[\eta']} \right),
\end{align}
and the tail-to-cone ratio is 
\begin{equation}
	\label{PhiTailToCone_v1}
	\bigg| \frac{\delta_1 C^{(\Phi|{\rm tail})i}_{\quad\qquad 0j0} }{\delta_1 C^{(\Phi|\gamma)i}_{\;\quad\quad 0j0}}\bigg| \sim  (H r_{\rm phy})(H\Delta t).
\end{equation}
For the case where the dS-like behavior is recovered asymptotically, each $r$ derivative generates an additional factor of $r/(\eta\eta')$. Moreover, we have $1\geq r/|\eta'| \gg \abs{\Vec{x}'}/|\eta'|$ by the causality. Therefore, the higher order multipole moments are suppressed by factor $\abs{\Vec{x}'}/\abs{\eta'}$. Thus,
\begin{align}
\label{LinearizedWeyl_Tail_Asymptotic}
\delta_1 C^{(\Phi|{\rm tail})i}_{\quad\qquad 0j0}  
& \approx -\frac{3\GN \delta w_0}{2a[\eta]} (\delta_{ij}-3\widehat{r}_i\widehat{r}_j)  \int_{-\infty}^{\eta_r} \dd\eta' \;\chubble(\eta)\chubble(\eta') \Delta[\eta,\eta',r] \left( M[\eta'] + \frac{\ddot{Q}_{\ell\ell}[\eta']}{2a^2[\eta']} \right) .
\end{align}
In this case, the tail-to-cone ratio is about
\begin{equation}
\label{PhiTailToCone_v2}
\bigg| \frac{\delta_1 C^{(\Phi|{\rm tail})i}_{\quad\qquad 0j0} }{\delta_1 C^{(\Phi|\gamma)i}_{\;\quad\quad 0j0}}\bigg| \sim  \delta w_0 \big(H \Delta t\big) \big( H r_{\rm phy}\big).
\end{equation}
Additionally, the scalar-to-tensor ratios for both cases are 
\begin{equation}
	\bigg| \frac{\delta_1 C^{(\Phi)i}{}_{0j0} }{\delta_1 C^{(g)i}{}_{0j0}}\bigg|\sim (H\tau_c)^2 \delta w.
\end{equation}

It is worthwhile to highlight, by comparing the spin$-2$ tidal forces in eq. \eqref{LinearizedWeyl_TensorNR} to the scalar ones in equations \eqref{LinearizedWeyl_NullFront}, \eqref{LinearizedWeyl_Tail_FromRest} and \eqref{LinearizedWeyl_Tail_Asymptotic}, we see they depend on distinct irreducible components of the acceleration of the mass quadrupoles -- the former depends on the `transverse-in-space' and `traceless' ones whereas the latter on their `trace'. This corroborates the appropriateness of their respective spin designations.

{\bf Polarization Patterns} \qquad To further understand the physical meaning of these ``trace-less" tidal forces, let us place two co-moving geodesic test masses separated by
\begin{align}
\label{DeviationVector}
\xi^0 \equiv 0
\qquad \text{and} \qquad
\vec{\xi}
\equiv \epsilon (\cos\vartheta \cdot \widehat{r} + \sin\vartheta \cdot \widehat{r}_\perp) ,
\end{align}
with $|\epsilon| \ll r$ and $\widehat{r}_\perp$ being the unit spatial vector orthogonal to $\widehat{r}$. The first order acceleration in eq. \eqref{TraceFreeTidalForce_1stOrder} at leading order in $\delta w$ and in the multi-pole expansion, according to eq. \eqref{LinearizedWeyl_NullFront}, is
\begin{align}
a^\mu \xi_\mu
\approx -\frac{\GN}{2a[\eta]r} \left( 3 \sin^2[\vartheta] - 2 \right) \chubble[\eta]\chubble[\eta_r] \sqrt{\delta w[\eta]\delta w[\eta_r]} \left( M[\eta_r] + \frac{\ddot{Q}_{\ell\ell}[\eta_r]}{2a^2[\eta_r]} \right) .
\end{align}
A circular ring of geodesic test masses would therefore experience {\it uniform} squeezing and stretching -- it would remain circular -- though the magnitude of the tidal forces do vary depending on the angle it makes with incident direction of the gravitational ration $\widehat{r}$. A similar scalar gravitational wave polarisation pattern was also obtained in the radiation-dominated universe \cite{Chu2021Gravitational}.

\section{Scalar Gravitational Radiation from Compact Binary System}
\label{sec: binary}
In this section, the linearised Weyl tensor results obtained in the previous section is applied to a compact binary system. The detail treatment of the quadrupolar gravitational radiation emission from a binary system in Minkowski background was studied in \cite{Peters1963Gravitational}, and the quadrupolar radiation emitted by a similar system in de Sitter spacetime was investigated in \cite{Bonga2017Power} and \cite{Hoque2019Quadrupolar}. For our purpose of studying far-field behaviour of gravitational waves, the binary system is modeled by the action of two point particles with mass $m_1$ and $m_2$,

\begin{align}
	\mathcal{S}_\text{astro} = -m_1 \int \dd t \sqrt{g_{\mu\nu} \frac{\dd z_1^\mu}{\dd t} \frac{\dd z_1^\mu}{\dd t}} -m_2 \int \dd t \sqrt{g_{\mu\nu} \frac{\dd z_2^\mu}{\dd t} \frac{\dd z_2^\mu}{\dd t}} .
\end{align}
Furthermore, we assume that the active duration of the binary system is much shorter than the cosmological scale, so that $a[\eta]$ can be approximated as a constant at the emission time. To simplify our calculation, we consider the non-relativistic Newtonian gravity limit, where the motion reduces to the Keplerian one. Let $e$ be the eccentricity of the orbit, $r_0$ the semi-major axis, and $\psi$ the angular coordinate. The spatial origin is placed at the center of mass, which is one focus of the orbit. The coordinate separation $d$ is then given by
\begin{equation}
	d = \frac{r_0 (1-e^2)}{1+e\cos\psi}.
\end{equation}
Without loss of generality, we assume that the motion of binary is confined on the $(1,2)$-plane. The separation vector is then given by 
\begin{equation}
	\Vec{d} = d ( \cos\psi,\sin\psi,0 ).
\end{equation}
The positions of two point masses are 
\begin{equation}
	\vec{d}_1 = \frac{m_2}{m_1+m_2} \vec{d}, \qquad \vec{d}_2 = -\frac{m_1}{m_1+m_2} \vec{d}.
\end{equation}
The energy-momentum tensor of the binary system is 
\begin{equation}
	T^{(a)}_{\mu\nu} = \frac{m_1}{\sqrt{-g}} \int \dd \tau \; \frac{\dd z_{1\mu}}{\dd \tau} \frac{\dd z_{1\nu}}{\dd\tau} \delta^{(4)}[x'-z_1] + \frac{m_2}{\sqrt{-g}} \int \dd \tau \; \frac{\dd z_{2\mu}}{\dd \tau} \frac{\dd z_{2\nu}}{\dd\tau} \delta^{(4)}[x'-z_2].
\end{equation}
In the Newtonian limit, its $00$-component is
\begin{equation}
	T^{(a)}_{00} \approx \frac{m_1}{a} \delta^{(3)}[\vec{x}'-\vec{d}_1] + \frac{m_2}{a} \delta^{(3)}[\vec{x}'-\vec{d}_2].
\end{equation}
Then the mass monopole is 
\begin{equation}
	M(\eta) \approx m_1 + m_2 \equiv m,
\end{equation}
and the mass quadrupole is 
\begin{equation}
	\label{BinaryQuadrupole}
	Q_{ij}(\eta) = a^2(\eta) \mu d^2 \begin{pmatrix}
		\cos^2\psi 			& \cos\psi \sin\psi & 0 \\
		\cos\psi \sin\psi 	& \sin^2\psi 		& 0 \\
		0 & 0 & 0
	\end{pmatrix},
\end{equation}
where $\mu \equiv m_1 m_2/(m_1+m_2)$.

\noindent{\bf Tensor contribution} \qquad Even though scalar emission is the primary goal of this paper, for completeness, we first present the tensor contribution to the Weyl tensor. To simplify the expression, we assume $e\ll 1$, and only extract the zeroth order in $e$. Inserting the fourth derivative of eq. \eqref{BinaryQuadrupole} into eq. \eqref{LinearizedWeyl_TensorNR} then returns the leading order tensor contribution
\begin{equation}\label{LinearizedWeyl_BinaryTensorGWs}
	\delta_1 C^{(g|\gamma)i}_{\;\quad\quad 0j0} \approx - \frac{8\GN^{5/3}\omega^{8/3}_a m^{2/3} \mu }{a[\eta]r}\left( \frac{1+\cos^2\theta}{2} \cos[2(\psi-\phi)]\epsilon^{+}_{ij} + \cos\theta \sin[2(\psi-\phi)]\epsilon^{\times}_{ij} \right),
\end{equation}
where $(r,\theta,\phi)$ represents the position of observer, $\omega_a$ is the angular frequency determined by the approximate Kepler's third law
\begin{equation}
	\omega^2_a = \frac{\GN m}{r^3_0},
\end{equation}
and $\epsilon^{+}$ and $\epsilon^\times$ are the polarisation tensors given by \cite{Poisson2014Gravity}
\begin{equation}
	\label{PolarizationsPM}
	\epsilon^+_{ij} = \widehat{\theta}_i\widehat{\theta}_j - \widehat{\phi}_i\widehat{\phi}_j, \qquad \epsilon^\times_{ij} = \widehat{\theta}_i\widehat{\phi}_j + \widehat{\phi}_i\widehat{\theta}_j,
\end{equation}
where $\widehat{\theta}$ and $\widehat{\phi}$ are unit vectors which are transverse to $\widehat{r}$.

\noindent{\bf Scalar contribution} \qquad The scalar contributions in eq. \eqref{LinearizedWeyl_NullFront} consist of the mass and pressure monopole terms. The mass monopole $M$ is approximately a constant over cosmological timescales $1/H_\star$; whereas $\ddot{Q}_{\ell\ell}$ scales as $M \cdot (r_s/\tau_c)^2$. Hence, the latter lies in a much higher frequency bandwidth than the former. Now,
\begin{equation}
	\ddot{Q}_{\ell\ell} = 2\mu a^2(\eta) (\GN m\omega_a)^{2/3} \frac{e(e+\cos\psi)}{1-e^2}.
\end{equation} 
Since we are focusing on the high frequency signal, we shall retain only the cosine term, leading us to the following scalar-induced electric Weyl tensor components:
\begin{align}
\label{LinearizedWeyl_BinaryScalarGWs}
\delta_1 C^{(\Phi|\gamma)i}_{\;\quad\quad 0j0}[\text{High freq.}]
& \approx -\frac{\GN^{5/3} \omega^{2/3}_a m^{2/3} \mu}{2a[\eta] r} \frac{e}{1-e^2} (\delta_{ij}-3\widehat{r}_i\widehat{r}_j)\chubble[\eta]\chubble[\eta_r] \sqrt{\delta w[\eta]\delta w[\eta_r]}  \cos\psi.
\end{align}
A circular ring of geodesic test masses would be squeezed and stretched at the same frequency as that of the binary system itself: compare the $2\psi$ in eq. \eqref{LinearizedWeyl_BinaryTensorGWs} versus the $\psi$ in eq. \eqref{LinearizedWeyl_BinaryScalarGWs} -- the factor of $2$ difference distinguishes spin$-2$ from spin$-0$. The eccentricity dependence also indicates circular orbits ($e=0$) do not produce scalar gravitational waves. Finally, taking the ratio of eq. \eqref{LinearizedWeyl_BinaryScalarGWs} to eq. \eqref{LinearizedWeyl_BinaryTensorGWs} informs us of the sobering fact that this new channel is suppressed relative to the usual tensor one by $e (\mathcal{H}^2/\omega_a^2) \delta w$.

\section{Discussion, Summary, and Outlook}
\label{sec: conclusion}
Even without direct coupling between the ordinary matter comprising astrophysical systems and any hypothetical fields that might be responsible for the accelerated expansion of the universe, we have argued in this paper that scalar, spin$-0$, gravitational waves may be generated because perturbations of these field(s) would mix with those of the gravitational metric. We demonstrated it in detail -- see eq. \eqref{Einstein_1stOrder} -- using a single canonical scalar field acting as Dark Energy; either by releasing it from ``rest" or allowing it to asymptote to a dS-like global minimum of a flat enough potential. The dominant null cone portion of the spin$-0$ signal is found to be directly sensitive to the Dark Energy equation-of-state; and, in the non-relativistic but high frequency limit, is proportional to the (Euclidean) spatial trace of the acceleration of the astrophysical system's mass quadrupole moments. By specializing to the physically important case of the compact binary system, we arrived at eq. \eqref{LinearizedWeyl_BinaryScalarGWs}, which describes the traceless tidal forces experienced by a small geodesic body in the far zone: these forces are directly proportional to $e/(1-e^2)$, where $e$ is the orbit's eccentricity. As a byproduct, we also showed that the dominant portion of the spin-2 signal coincides with the de Sitter one.

To be sure, \cite{Garoffolo2020Gravitational,Dalang2021Scalar,Kubota2023Propagation} have -- motivated by Dark Energy -- studied the possibility of scalar gravitational waves in our current universe. They even examined more general scalar field models than ours; specifically, the subset of Horndeski theory \cite{Horndeski1974Second} that yields unit speed propagation of the tensor modes, since the latter has been well constrained recently. However, all of them chose to ignore the actual source of the gravitational radiation; instead focusing on the propagation properties by using JWKB techniques. Additionally, they also chose to `gauge-fix' to perform their calculations; and employed particular ansatz to distinguish ``scalar" from ``tensor" modes. As far as we are aware, ours is the first explicit gauge-invariant calculation of spin$-0$ gravitational radiation generated by a hypothetical compact binary system in a Dark Energy dominated universe.

On the other hand, the real universe is made up of approximately 70\% Dark Energy, 25\% Dark Matter, and 5\% ordinary matter. To make a concrete prediction of scalar gravitational waves of relevance to our cosmology, therefore, we need to extend our work to include (at the very least) Dark Matter, since this paper only accounted for a single Dark Energy scalar. Further understanding of cosmological scalar gravitational radiation would also be achieved if we can calculate the counterpart of the pseudo-stress tensor of gravitational waves in Minkowski spacetime. What are the energy and angular momentum loss (quadrupole) formulas for both scalar and tensor gravitational waves in cosmology? These would require a second order perturbative analysis, as opposed to the first order one we carried out here. Finally, even for such a simple canonical scalar field undergoing ``low energy slow-roll inflation", the tail signal of these scalar tidal forces turned out to be difficult to deal with due to the multiple time integrals appearing in the final expressions -- see equations \eqref{LinearizedWeyl_Tail_FromRest} and \eqref{LinearizedWeyl_Tail_Asymptotic}. Perhaps more techniques or numerical analysis are needed to extract deeper physical insight into their properties.

\section{Acknowledgments}
LYC thanks Jen-Yu Lo for discussion about the dynamical system analysis. YZC and YWL were supported by the National Science and Technology Council of the R.O.C. under Project No.~NSTC 112-2811-M-008-006 and 111-2112-M-008-003.

\newpage

\bibliographystyle{abbrv}
\bibliography{ScalarGWsFromDE}

\end{document}